\documentclass[10pt,a4paper]{article}
\usepackage[T1]{fontenc}
\usepackage[utf8]{inputenc}
\usepackage{amsmath,amssymb,bm}
\usepackage{newtxtext,newtxmath}
\usepackage[a4paper,left=18mm,right=18mm,top=19mm,bottom=21mm,columnsep=6.5mm]{geometry}
\usepackage{graphicx}
\usepackage{latexml}
\usepackage{booktabs,multirow}
\usepackage{microtype}
\usepackage[super,sort&compress,numbers]{natbib}
\usepackage{caption}
\usepackage{stfloats}
\usepackage{balance}
\usepackage{multicol}
\usepackage{titlesec}
\usepackage{xurl}
\usepackage[hidelinks]{hyperref}
\hypersetup{pdftitle={PEACE: Covariant learning of nonadiabatic manifolds with parity-resolved Hamiltonians},pdfauthor={Rongzhi Gao et al.}}
\renewcommand{\figurename}{Fig.}
\titleformat{\section}{\large\bfseries}{\thesection}{0.6em}{}
\titlespacing*{\section}{0pt}{10pt plus 2pt minus 1pt}{5pt}
\titleformat{\paragraph}[runin]{\normalsize\bfseries}{}{0pt}{}
\titlespacing*{\paragraph}{0pt}{6pt plus 1pt minus 1pt}{0.5em}
\newif\ifincludesi
\includesitrue

\newsavebox{\SIpagebox}
\newcommand{\SIpageblock}[2]{%
  \iflatexml
    #2%
  \else
    \sbox{\SIpagebox}{\begin{minipage}[t]{\textwidth}#2\end{minipage}}%
    \noindent\usebox{\SIpagebox}\par
  \fi
}

\newcommand{\PEACEgraphic}[2]{%
  \iflatexml%
    \includegraphics[width=174mm]{figures/#2.png}%
  \else%
    \includegraphics[#1]{figures/#2.png}%
  \fi%
}%
\newcommand{\SIfigure}[4]{%
  \iflatexml%
    \begin{figure}%
    \centering%
    \PEACEgraphic{#1}{#2}%
    \caption{#3}%
    \label{#4}%
    \end{figure}%
  \else%
    {\centering%
    \PEACEgraphic{#1}{#2}\par}%
    \captionof{figure}{#3}%
    \label{#4}%
  \fi%
}%
\newcommand{\SIsidefigure}[4]{%
  \iflatexml%
    \SIfigure{#1}{#2}{#3}{#4}%
  \else%
    \noindent\begin{minipage}[c]{0.55\textwidth}%
    \centering%
    \PEACEgraphic{#1}{#2}%
    \end{minipage}\hfill\begin{minipage}[c]{0.41\textwidth}%
    \captionof{figure}{#3}%
    \label{#4}%
    \end{minipage}%
  \fi%
}%
\newcommand{\SIhalffigure}[4]{%
  \iflatexml%
    \SIfigure{#1}{#2}{#3}{#4}%
  \else%
    \begin{minipage}[t]{0.48\textwidth}%
    \SIfigure{#1}{#2}{#3}{#4}%
    \end{minipage}%
  \fi%
}%
\newcommand{\SItable}[3]{%
  \iflatexml%
    \begin{table}%
    \centering%
    \caption{#1}%
    \label{#2}%
    #3%
    \end{table}%
  \else%
    \begin{minipage}[t]{\textwidth}%
    \centering%
    \captionof{table}{#1}%
    \label{#2}%
    #3%
    \end{minipage}%
  \fi%
}%

\begin{document}
\twocolumn[%
\begin{@twocolumnfalse}
\begin{center}
{\LARGE\bfseries PEACE: Covariant learning of nonadiabatic manifolds\\[2pt]
with parity-resolved Hamiltonians\par}
\vspace{8pt}
{\large Rongzhi Gao\textsuperscript{1,2,*},
Shuguang Chen\textsuperscript{2,3},
Yang Zhou\textsuperscript{1},
GuanHua Chen\textsuperscript{1,2,*},\\[3pt]
Ziyang Hu\textsuperscript{1,2,*} and
ChiYung Yam\textsuperscript{2,4,*}\par}
\vspace{6pt}
{\small
\textsuperscript{1}Department of Chemistry, The University of Hong Kong, Pok Fu Lam, Hong Kong SAR, China\\
\textsuperscript{2}Hong Kong Quantum AI Lab, Pak Shek Kok, Hong Kong SAR, China\\
\textsuperscript{3}MattVerse Limited, Pak Shek Kok, Hong Kong SAR, China\\
\textsuperscript{4}Shenzhen Institute for Advanced Study, University of Electronic Science and Technology of China, Shenzhen, China\par}
\vspace{5pt}
{\small \textsuperscript{*}Corresponding authors:
\href{mailto:rzgao@yangtze.hku.hk}{rzgao@yangtze.hku.hk};
\href{mailto:ghc@everest.hku.hk}{ghc@everest.hku.hk};\\
\href{mailto:hzy@yangtze.hku.hk}{hzy@yangtze.hku.hk};
\href{mailto:yamcy@uestc.edu.cn}{yamcy@uestc.edu.cn}\par}
\end{center}
\vspace{3pt}
\begin{center}\bfseries Abstract\end{center}
\noindent

Nonadiabatic molecular dynamics provides mechanistic insight into light-driven processes and informs the design of molecules and materials for solar energy conversion, photocatalysis and photo switching. Accurately describing these processes requires a representation that respects electronic symmetry and consistently relates energies to interstate couplings. Here we introduce PEACE, which combines a parity-equivariant latent Hamiltonian with a learned electronic connection. Controlled ablations reveal the complementary roles of symmetry-allowed state mixing and electronic-frame variation in reproducing crossing structures and relaxation dynamics. PEACE closely reproduces excited-state population dynamics from first-principles simulations, while its extension to spin--orbit coupling enables simulations of intersystem crossing. These results demonstrate that a more complete incorporation of the underlying physics into learned electronic representations leads to more accurate predictions of nonadiabatic dynamics.

\par\vspace{12pt}\end{@twocolumnfalse}]

\section*{Introduction}

Nonadiabatic molecular dynamics describes the coupled evolution of electrons and nuclei, providing an atomistic account of photochemical reactions\cite{main1,main2,main3,main4,main5}. Its computational cost arises from repeatedly evaluating multiple electronic energies, forces and interstate couplings along trajectory ensembles. The success of equivariant machine learning potentials in ground-state simulations\cite{main6,main7,main8,main9,main10,main11} has motivated their extension to excited states\cite{main12,main13,main14,main15,main16,main17,main18}. Methods such as SPaiNN\cite{main19}, SchNarc\cite{main20}, exciting DeepMD\cite{main21} and the diabatic artificial neural network (DANN)\cite{main22} have enabled efficient simulations of molecular photo dynamics.

Learning nonadiabatic couplings, however, presents challenges beyond predicting energy surfaces. These couplings depend on electronic-state phases, can diverge as energy gaps close and may exhibit double-valued behaviour associated with geometric phases. Energy-gap weighting removes the explicit inverse-gap factor, but the resulting coupling field is generally non-conservative. SchNarc addresses phase ambiguity through phase-invariant training\cite{main20}, yet its scalar-gradient coupling formulation imposes a conservative-field approximation and can suppress symmetry-allowed out-of-plane components at planar geometries. SPaiNN supports direct vectorial coupling prediction through equivariant readouts\cite{main19,main23,main24}, while Exciting DeePMD learns phase-invariant coupling dyads\cite{main21}. These more flexible representations nevertheless predict energies and couplings through separate readouts, without explicitly connecting them through a shared electronic Hamiltonian.

Learning an electronic Hamiltonian provides such a connection, as demonstrated by DANN and symmetry-adapted quasi-diabatic models\cite{main22}. Two physical considerations remain central to this representation. First, Hamiltonian matrix elements must respect electronic-state symmetry: unlike energies, off-diagonal interactions need not be invariant under reflection\cite{main25}. Second, derivative couplings contain contributions from both Hamiltonian eigenvector rotation and changes in the underlying electronic basis\cite{main26,main27}. Since a strictly diabatic basis is not generally available for a finite electronic subspace of polyatomic molecules, these residual couplings can influence dynamics. These considerations motivate treating symmetry-dependent state mixing and electronic-frame variation as complementary parts of the learned representation.

Here we introduce \textbf{P}arity-\textbf{E}quivariant \textbf{A}nd \textbf{C}ovariant \textbf{E}xcited-states learning (PEACE) framework, an equivariant framework that combines a latent Hamiltonian with a learned electronic connection. Explicit parity constraints determine how Hamiltonian blocks transform under reflection and permit symmetry-allowed mixing between latent states of opposite parity. The connection models changes in the latent electronic frame as the nuclei move. Hamiltonian eigenvalues provide adiabatic energies, while a common covariant derivative yields conservative forces and energy-gap-weighted nonadiabatic couplings. We examine the complementary roles of these components through controlled ablations, benchmark the resulting alkene photo dynamics and extend the framework to dynamics involving spin--orbit coupling. Across the tested systems, PEACE achieves accurate predictions of energies, forces and interstate couplings and captures the key features of excited-state population dynamics. Together with the controlled ablations, these results demonstrate the importance of incorporating physical structure into the learned electronic representation, including electronic symmetry and changes in the electronic frame, for reliable nonadiabatic dynamics.

\section*{Results}

\paragraph{The architecture of PEACE.}

We developed PEACE to learn a consistent electronic representation that connects adiabatic energy surfaces with the couplings governing nonadiabatic transitions (\textbf{Fig. 1}). Starting from atomic species and nuclear coordinates, a shared \emph{O}(3)-equivariant tensor-product network\cite{main6,main28} constructs molecular features with defined transformation properties under rotations and reflections. Independently parameterized equivariant attention branches\cite{main10} use these features to predict a real symmetric latent Hamiltonian \textbf{H}, and a real antisymmetric electronic connection \(\mathbf{B}_{\mu}\). These quantities jointly describe electronic-state interactions and the contribution of the changing latent frame to derivative couplings.

Reflection symmetry is incorporated by assigning fixed parity labels to the latent basis states. Hamiltonian matrix elements connecting states of equal parity are constructed from even-scalar features, whereas those connecting opposite-parity states use odd-pseudoscalar features. The latter change sign between mirror-related geometries and vanish at geometries invariant under reflection. Their derivatives along symmetry-breaking coordinates can nevertheless be nonzero, allowing the model to represent the onset of electronic-state mixing as the molecule moves away from a symmetric configuration. This structure provides a symmetry-adapted description of local gap opening near electronic crossings while retaining nonlinear geometric dependence\cite{main29}. The adiabatic states emerge as geometry-dependent combinations of the latent basis states.

\begin{figure*}[!t]
\centering
\PEACEgraphic{width=\textwidth,height=0.72\textheight,keepaspectratio}{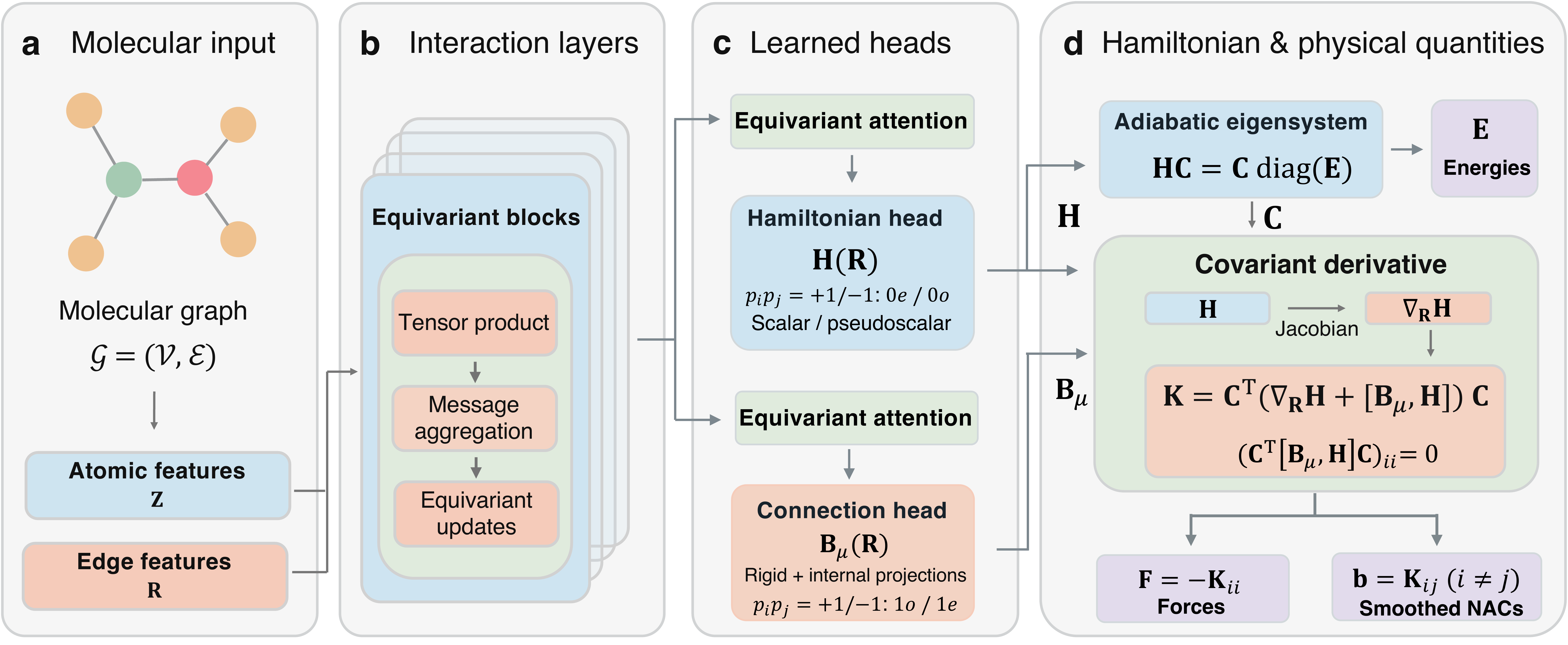}
\caption{\textbf{The overall architecture of PEACE framework.} \textbf{a}, The atomic species \textbf{Z} and Cartesian coordinates \textbf{R} define a molecular graph \(\mathcal{G\  = \ (V,\ E)}\). Nodes \(\mathcal{V}\) are represented by one-hot encodings of atomic species, while edges \(\mathcal{E}\) are described by radial basis functions of interatomic distances and spherical harmonics of interatomic directions. \textbf{b,} A shared \emph{O}(3)-equivariant tensor-product encoder combines message aggregation and equivariant updates. \textbf{c}, Two independently parameterized equivariant self-attention branches feed the Hamiltonian and connection heads. In the selected electronic basis, \(p_{i} \in \{ + 1,\  - 1\}\) are latent-basis parity labels, not labels fixed to energy-sorted adiabatic states. Same-parity diabatic Hamiltonian (\textbf{H}) entries use even scalar 0\emph{e} features, and opposite-parity entries use odd pseudoscalar 0\emph{o} features. Pseudoscalars construct Hamiltonian entries and their coordinate derivatives enter the physical readout. Connection same or opposite-parity pairs use polar 1\emph{o} or axial 1\emph{e} channels, respectively. \(\mathbf{B}_{\mu}\) is antisymmetric in electronic indices and combines directly learned rigid and internal motion components using the implemented regularized projections. \textbf{d}, The adiabatic energies \textbf{E} are eigenvalues of \textbf{H}, while the forces \textbf{F} and smoothed non-adiabatic couplings (NACs) \textbf{b} are calculated through covariant derivatives.}
\label{fig:1}
\end{figure*}

The learned connection accounts for an additional source of geometric dependence: changes in the latent electronic frame as the nuclei move. Adiabatic derivative couplings therefore contain contributions from both the coordinate dependence of the Hamiltonian eigenvectors and the learned connection. To respect reflection symmetry, connection blocks between equal-parity and opposite-parity latent states are constructed from polar-vector and axial-vector features, respectively. The connection also combines learned rigid-motion and internal-motion vector fields through regularized projections, allowing both types of nuclear motion to contribute to the coupling representation.

Adiabatic energies are obtained by diagonalizing \textbf{H}. Forces and smoothed nonadiabatic couplings (SNACs) follow from the covariant derivative transformed into the adiabatic basis,

\begin{equation}
\mathbf{K}=\mathbf{C}^{T}\bigl(\nabla_{\mathbf R}\mathbf H+[\mathbf B_\mu,\mathbf H]\bigr)\mathbf C.
\label{eq:1}
\end{equation}

where the columns of \textbf{C} are the Hamiltonian eigenvectors. The negative diagonal elements of \textbf{K} give the force components, while its off-diagonal elements give the energy-gap-weighted couplings. Because the connection commutator has zero diagonal in the adiabatic basis, the forces retain the conservative relation \(F_{i\mu} = - \partial_{\mu}E_{i}\). The connection thus expands the coupling representation while preserving consistency between energies and forces. PEACE is trained jointly on energies, forces and SNACs, linking the supervision of these observables through the shared electronic representation. Numerical transformation tests yielded relative deviations of order 10\textsuperscript{-15} under rotations, reflections and translations, consistent with the imposed symmetry equivariant constraints (\textbf{Supplementary Fig. 15}).

\begin{figure*}[!t]
\centering
\PEACEgraphic{width=\textwidth,height=0.72\textheight,keepaspectratio}{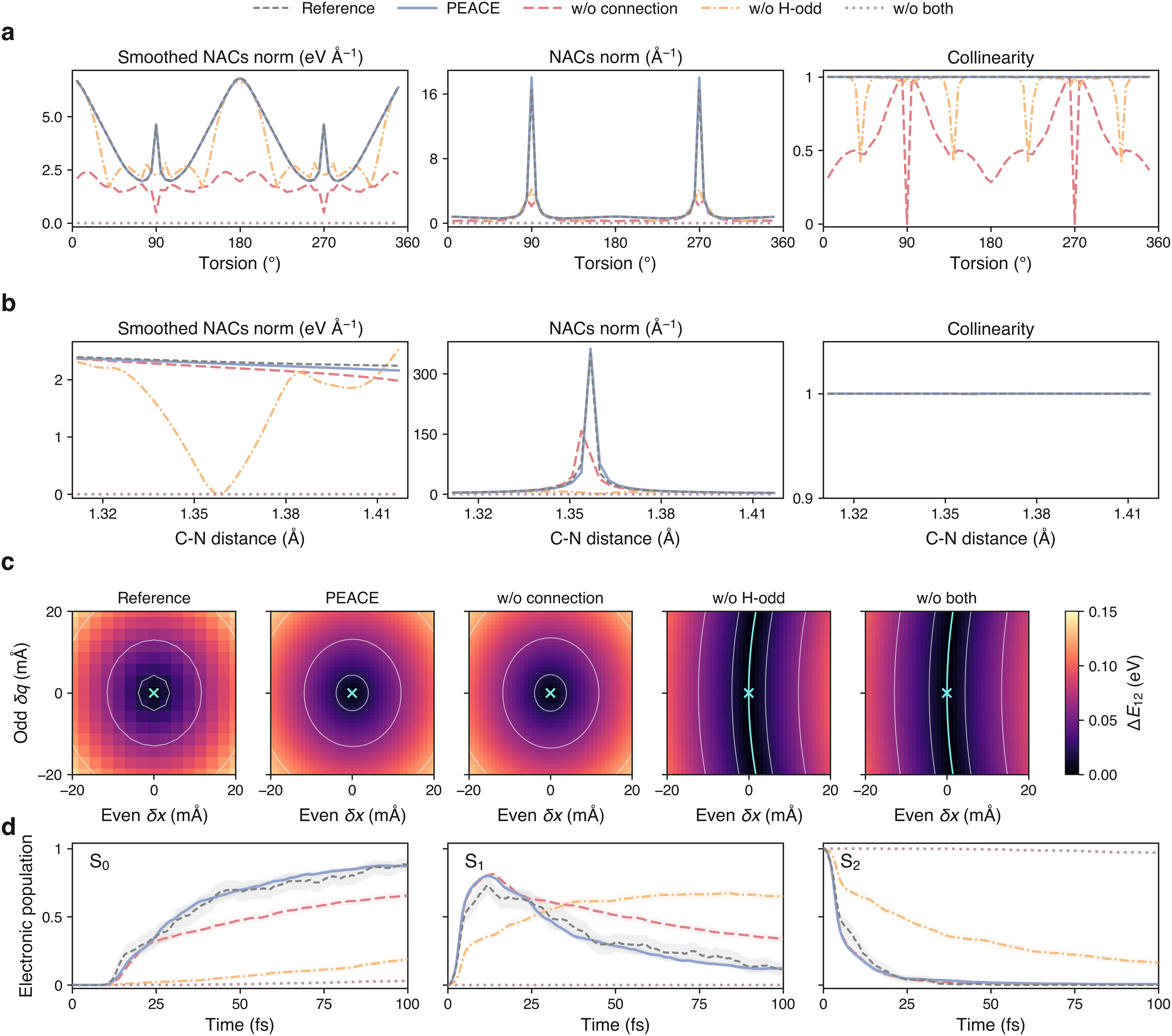}
\caption{\textbf{Couplings, crossing topology and dynamics of CH\textsubscript{2}NH\textsubscript{2}\textsuperscript{+}}. Curves compare PEACE with models lacking the connection, the opposite-parity Hamiltonian block or both. \textbf{a}, \textbf{b}, smoothed nonadiabatic coupling (NAC) and raw NAC norms, and phase-insensitive absolute collinearity of the predicted and reference coupling vectors, along an S\textsubscript{0}/S\textsubscript{1} torsional scan and an S\textsubscript{1}/S\textsubscript{2} C--N stretching scan. Undefined directions for vanishing predicted couplings are omitted. \textbf{c}, S\textsubscript{1}/S\textsubscript{2} energy-gap maps in common even and odd rigid-motion-free displacement directions; the reference and each model are centred on their own mirror-plane crossing. The MR-CISD map uses a 15 × 15 grid, while model maps use 41 × 41 grids. White contours mark gaps of 0.02, 0.06 and 0.12 eV. \textbf{d}, Mean electronic populations from trajectories initialized in S\textsubscript{2}. Shading shows pointwise 95\% intervals from trajectory-bootstrap resamples. The reference and four model ensembles contain 145, 919, 926, 901 and 960 independently retained trajectories in legend order with maximum energy drift less than 0.25 eV.}
\label{fig:2}
\end{figure*}

\paragraph{Ablation of odd-parity and electronic connection.}

We examined the contributions of the learned connection and opposite-parity Hamiltonian blocks through controlled ablations on the methaniminium cation molecule (CH\textsubscript{2}NH\textsubscript{2}\textsuperscript{+})\cite{main12,main20}. Full PEACE was compared with three variants excluding the connection, the opposite-parity Hamiltonian blocks, or both. All models shared the data split and initialization, and their checkpoints were selected using the validation loss (\textbf{Supplementary Fig. 1}). We assessed both prediction errors on 1,250 test geometries and the resulting crossing structures and relaxation dynamics.

Removing the connection restricts derivative couplings to the contribution arising from the coordinate dependence of the Hamiltonian eigenvectors. This ablation increased the SNAC mean absolute error (MAE) from 0.117 to 0.410 eV Å\textsuperscript{-1} (\textbf{Supplementary Fig. 2}). Coupling directions deteriorated along the torsional scan, although the localized gap minimum remained in the two-dimensional map around the S\textsubscript{1}/S\textsubscript{2} crossing (\textbf{Fig. 2 a} and \textbf{c}). Surface-hopping simulations also showed slower depletion of S\textsubscript{1} (\textbf{Fig. 2d}). These results indicate that the explicit connection improves the representation of derivative couplings beyond the information supplied by Hamiltonian eigenvector variations in the tested model.

Retaining the connection while removing the opposite-parity Hamiltonian blocks led to a different failure. This model reached a low training loss, yet its test-set SNAC MAE increased to 0.257 eV Å\textsuperscript{-1}, and the localized gap minimum was replaced by an extended low-gap valley (\textbf{Supplementary Figs. 1} and \textbf{2} and \textbf{Fig. 2c}). Distorted coupling profiles and persistent excited-state populations accompanied this change. Thus, a good fit to the sampled training observables did not ensure recovery of the reference crossing structure. The distinction follows from the roles of the two components: the connection contributes to derivative couplings, while the energy-gap landscape is determined by the Hamiltonian. For a fixed Hamiltonian, learning the connection cannot directly restore the missing interactions between opposite-parity sectors. The H-odd blocks provide these symmetry-allowed interactions and enable state mixing along symmetry-breaking coordinates.

Removing both components increased the SNAC MAE to 0.687 eV Å\textsuperscript{-1} and strongly suppressed relaxation from S\textsubscript{2}, despite retaining an energy MAE of 0.033 eV. Full PEACE instead captured the initial S\textsubscript{2} depopulation, transient S\textsubscript{1} accumulation and subsequent S\textsubscript{0} growth. Together, these ablations show that low training loss and small energy errors provide an incomplete assessment of nonadiabatic models. When compared with other models\cite{main19,main20,main21}, PEACE also achieve the state-of-the-art (SOTA) performance in both static and dynamics process (\textbf{Supplementary Figs. 3-5}). Agreement with the reference crossing structure and population dynamics benefits from combining symmetry-allowed Hamiltonian mixing with an explicit representation of the latent-frame contribution to derivative couplings.

\paragraph{Benchmarking alkene photo dynamics.}

We next evaluated PEACE on C\textsubscript{2}H\textsubscript{4}, C\textsubscript{3}H\textsubscript{6}, and C\textsubscript{4}H\textsubscript{8} using SA(3)-\allowbreak{}CASSCF(2,2)/\allowbreak{}cc-pVDZ reference calculations, with SchNet\cite{main23}- and PaiNN\cite{main24}-based SPaiNN\cite{main19} models included as comparative benchmarks. To improve coverage of the S\textsubscript{1}/S\textsubscript{0} transition regions, we expanded the original datasets\cite{main19} and retrained the SPaiNN models using the same batch size and number of training epochs as PEACE.

Among the models evaluated, PEACE achieved the SOTA overall performance across the three alkene benchmarks with energy mean absolute errors (MAEs) of 0.008--0.022 eV and force MAEs of 0.036--0.054 eV Å\textsuperscript{-1} (\textbf{Supplementary Table 1}). The corresponding SNAC MAEs were 0.066, 0.041 and 0.069 eV Å\textsuperscript{-1} for C\textsubscript{2}H\textsubscript{4}, C\textsubscript{3}H\textsubscript{6}, and C\textsubscript{4}H\textsubscript{8}, respectively, supporting accurate prediction of both the energy surfaces and interstate couplings (\textbf{Supplementary Figs. 6, 8} and \textbf{10}).

This agreement extended to the excited-state population dynamics (\textbf{Fig. 3 a-c}). Effective S\textsubscript{1} decay time constants obtained from a common one-step fit over 0--100 fs was 68.5, 51.4 and 67.8 fs for PEACE, closely matching the reference values of 68.8, 53.3 and 66.9 fs, respectively (\textbf{Supplementary Figs. 7, 9} and \textbf{11}). PEACE maintained close agreement with the reference population evolution across all three alkenes, with fitted decay times differing by at most 1.9 fs.

PEACE also substantially reduced the per-geometry evaluation time under the tested hardware configurations (\textbf{Supplementary Fig. 14}). FP64 PEACE inference required 7.66--12.44 ms on a single NVIDIA H100 GPU, compared with 6.77--13.39 s for the CASSCF workflow using six CPU workers on an Intel Xeon Platinum 8480C system. These hardware-specific wall-clock ratios of \textasciitilde1,000 demonstrate the practical computational advantage of PEACE for ensemble nonadiabatic dynamics.

\begin{figure*}[!t]
\centering
\PEACEgraphic{width=0.88\textwidth,height=0.72\textheight,keepaspectratio}{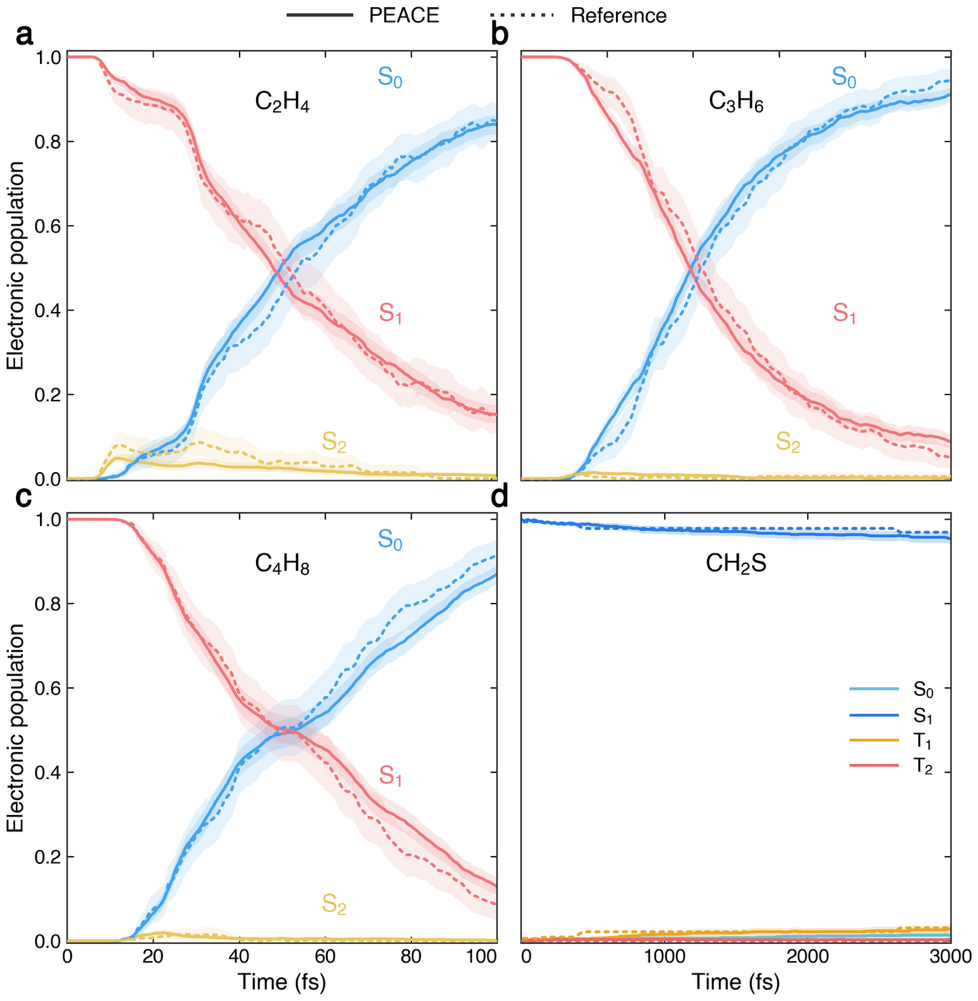}
\caption{\textbf{Ensemble electronic populations in four molecules}. Solid curves show PEACE and dotted curves show the reference. The maximum energy drift of all trajectories is less than 0.25 eV. \textbf{a}-\textbf{c} C\textsubscript{2}H\textsubscript{4}, C\textsubscript{3}H\textsubscript{6}, and C\textsubscript{4}H\textsubscript{8} over 100 fs. The reference ensemble contains 200 trajectories for each system; PEACE ensembles contain 919, 906 and 920, respectively. Shading is the archived pointwise bootstrap interval. \textbf{d}, CH\textsubscript{2}S over 3 ps, with 954 PEACE trajectories and a pointwise interval standard errors. The reference curves were extracted from the reference\cite{main20}.}
\label{fig:3}
\end{figure*}

\paragraph{Spin--orbit couplings and intersystem crossing.}

We extended PEACE to describe spin--orbit couplings (SOCs)\cite{main30,main31} by retaining separate Hamiltonian and connection representations for the singlet and triplet manifolds and adding an equivariant SOC readout. This readout predicts Cartesian coupling vectors with the required rotation and reflection symmetries, which are combined with fixed spin-angular coefficients to construct a complex Hermitian SOC matrix (\textbf{Methods}). We evaluated the resulting model on CH\textsubscript{2}S. On the test set, the energy, force and SNAC MAEs were 1.113 meV, 4.941 meV Å\textsuperscript{-1} and 0.933 meV Å\textsuperscript{-1}, respectively. Complex SOC matrix elements were predicted with an MAE of 0.265 cm\textsuperscript{-1} and an RMSE of 0.397 cm\textsuperscript{-1} (\textbf{Supplementary Fig. 12}). Along the C--S bond-length scan, PEACE also reproduced the reference energy profiles, selected SOC components and the decrease in the singlet and triplet SNAC norms with increasing bond length (\textbf{Supplementary Fig. 13}). Including both nonadiabatic and spin--orbit couplings enabled surface-hopping propagation over 3 ps from S\textsubscript{1}-initialized Wigner conditions (\textbf{Fig. 3d}). At 3 ps, the retained ensemble had an S\textsubscript{1} population of 95.4\% and a T\textsubscript{1} population of 2.8\%, indicating weak population transfer to the triplet manifold over this interval. The predominantly S\textsubscript{1} population and small triplet contribution are consistent with the published CASSCF reference\cite{main20}. This dynamical comparison concerns population trends, complementing the direct static evaluation of the predicted couplings.

\section*{Discussion}

Our PEACE framework demonstrates the value of incorporating electronic symmetry and frame variation into a common representation of excited-state properties. The controlled ablations reveal complementary contributions from the opposite-parity Hamiltonian blocks and the electronic connection to crossing structure and coupling predictions. Their effects on population dynamics show that low training loss alone is insufficient to establish physical fidelity, motivating the assessment of local crossing behaviour and ensemble relaxation alongside static prediction errors.

A promising direction is to extend PEACE to condensed environments\cite{main32}, including solvated molecules and molecular assemblies. Incorporating suitable descriptions of environmental and intermolecular interactions could enable studies of how solvation and aggregation influence excited-state relaxation, broadening the framework's relevance to photochemistry in solution and molecular materials. Active learning\cite{main33} could support these extensions by selecting informative configurations while controlling the cost of reference calculations. Additional electronic-structure information, such as state overlaps\cite{main34}, could also be explored as optional training supervision, with potential gains in accuracy or data efficiency weighed against the added cost of data generation and training. These developments would aim to preserve efficient inference while extending the range of photochemical and photophysical processes accessible to PEACE.

\section*{Methods}

\paragraph{Parity-resolved and covariant representations.}

For nuclear \(N_{\text{atom}}\) atoms' coordinates \(\mathbf{R} \in \mathbb{R}^{3N_{\text{atom}}}\), PEACE predicts a real symmetric latent Hamiltonian \(\mathbf{H}(\mathbf{R})\) and a real antisymmetric connection \(\mathbf{B}_{\mu}\left( \mathbf{R} \right)\), where \(\mu = (A,\alpha)\) denotes Cartesian component \(\alpha\) of atom \(A\). In an orthonormal latent frame\cite{main26} \(\left\{ |\chi_{a}\rangle \right\}\), \(B_{\mu,ab}\) formally represents \(\left\langle \chi_{a}|\partial_{\mu}\chi_{b} \right\rangle\). We define the covariant Hamiltonian derivative as,

\begin{equation}
\mathbf G_\mu=\partial_\mu\mathbf H+[\mathbf B_\mu,\mathbf H]
\label{eq:2}
\end{equation}

\begin{equation}
[\mathbf B_\mu,\mathbf H]=\mathbf B_\mu\mathbf H-\mathbf H\mathbf B_\mu.
\label{eq:3}
\end{equation}

Diagonalization yields \(\mathbf{HC = CE}\), \(\mathbf{C}^{T}\mathbf{C = I}\)\textbf{,} where the columns of \(\mathbf{C}\) are latent-Hamiltonian eigenvectors and \(\mathbf{E}\) contains their adiabatic energies. Away from exact degeneracies, the adiabatic derivative coupling and covariant derivative satisfy,

\begin{equation}
\mathbf d_\mu=\mathbf C^T\partial_\mu\mathbf C+\mathbf C^T\mathbf B_\mu\mathbf C
\label{eq:4}
\end{equation}

\begin{equation}
\mathbf K_\mu=\mathbf C^T\mathbf G_\mu\mathbf C=\partial_\mu\mathbf E+[\mathbf d_\mu,\mathbf E].
\label{eq:5}
\end{equation}

Consequently, the force on state \(i\) and the smoothed nonadiabatic coupling between states \(i\) and \(j\) are obtained from the same derivative,

\begin{equation}
F_{i\mu}=-(\mathbf K_\mu)_{ii},
\label{eq:6}
\end{equation}

\begin{equation}
\begin{aligned} b_{ij,\mu}&=(\mathbf K_\mu)_{ij}\\ &=(E_j-E_i)d_{ij,\mu}\quad(i\ne j). \end{aligned}
\label{eq:7}
\end{equation}

Here \(d_{ij,\mu} = \left\langle \psi_{i}|\partial_{\mu}\psi_{j} \right\rangle\) is the nonadiabatic coupling, with \(|\psi_{i}\rangle = \sum_{a}^{}|\chi_{a}\rangle C_{ai}\).

The connection combines learned the unprojected, atom-resolved Cartesian rigid-body \(\mathbf{X}_{ab}^{rig}\) and internal \(\mathbf{X}_{ab}^{int}\) vector fields,

\begin{equation}
\begin{aligned}\mathbf B_{ab}={}&\alpha_{\mathrm{rig}}\boldsymbol{\Pi}_{\mathrm{rig}}^{(\epsilon)}\mathbf X_{ab}^{\mathrm{rig}}\\ &+\alpha_{\mathrm{int}}\boldsymbol{\Pi}_{\mathrm{int}}^{(\epsilon)}\mathbf X_{ab}^{\mathrm{int}},\end{aligned}
\label{eq:8}
\end{equation}

where \(\mathbf{B}_{ab}\) collects all \(3N_{atom}\) coordinate components, \(\mathbf{\Pi}_{rig}^{(\epsilon)}\) is a regularized rigid-motion projection, \(\mathbf{\Pi}_{int}^{(\epsilon)} = \mathbf{I} - \mathbf{\Pi}_{rig}^{(\epsilon)}\), and \(\alpha_{rig}\) and \(\alpha_{int}\ \)are fixed scale factors.

Each latent basis state has a fixed parity \(p_{a} \in \{ + 1,\  - 1\}\), collected in \(\mathbf{P} = diag\left( p_{1},\ldots,p_{N_{latent}} \right)\). For an orthogonal spatial transformation \(\mathbf{Q}\), the latent representation is \(\rho\left( \mathbf{Q} \right) = \mathbf{I}\) when \(\det\mathbf{Q} = + 1\) and \(\rho\left( \mathbf{Q} \right) = \mathbf{P}\) when \(\det\mathbf{Q} = - 1\). The predictions obey,

\begin{equation}
\mathbf H(\mathbf Q\mathbf R)=\rho(\mathbf Q)\mathbf H(\mathbf R)\rho(\mathbf Q)^T,
\label{eq:9}
\end{equation}

\begin{equation}
\begin{aligned}\mathbf B_{A\alpha}(\mathbf Q\mathbf R)={}&\sum_\beta Q_{\alpha\beta}\rho(\mathbf Q)\\ &{}\times\mathbf B_{A\beta}(\mathbf R)\rho(\mathbf Q)^T.\end{aligned}
\label{eq:10}
\end{equation}

Equal-parity pairs therefore use even-scalar \((0e)\) Hamiltonian channels and polar-vector (\(1o\)) connection channels; opposite-parity pairs use odd-scalar (\(0o\)) and axial-vector (\(1e\)) channels. The \(0o\) features arise from geometric pseudoscalars and equivariant convolutions. This construction applies at both symmetric and asymmetric geometries. If a reflection \(g\), together with any required permutation of identical atoms, leaves \(\mathbf{R}_{s}\) unchanged, then

\begin{equation}
H_{ab}(\mathbf R_s)=p_ap_bH_{ab}(\mathbf R_s).
\label{eq:11}
\end{equation}

Thus \(H_{ab}\left( \mathbf{R}_{s} \right) = 0\) for opposite-parity latent states. For a symmetry-odd distortion \(q\mathbf{u}\), such that \(g\left( \mathbf{R}_{s} + q\mathbf{u} \right) = \mathbf{R}_{s} - q\mathbf{u}\), the corresponding matrix element obeys,

\begin{equation}
\begin{aligned}H_{ab}(\mathbf R_s+q\mathbf u)&=-H_{ab}(\mathbf R_s-q\mathbf u)\\ &=qA_{ab}+O(q^3),\\ &\hspace{1em}p_ap_b=-1.\end{aligned}
\label{eq:12}
\end{equation}

The first-order term describes the local onset of coupling; PEACE retains nonlinear geometric dependence away from the symmetric point. At a general asymmetric geometry, opposite-parity latent sectors can therefore mix.

\paragraph{Spin-orbit couplings.}

Let \(n_{S}\) and \(n_{T}\) be the numbers of singlet and triplet spin-free roots. Expanding each triplet over \(M_{S} = - 1,0, + 1\) gives \(D = n_{S} + 3n_{T}\) spin components. The complex Hermitian Hamiltonian is assembled as,

\begin{equation}
\begin{aligned}\mathbf H_{\mathrm{full}}={}&\operatorname{diag}(\mathbf H_S,\mathbf H_T,\mathbf H_T,\mathbf H_T)\\ &+\mathbf V_{\mathrm{SO}},\end{aligned}
\label{eq:13}
\end{equation}

\begin{equation}
\mathbf V_{\mathrm{SO}}^\dagger=\mathbf V_{\mathrm{SO}},
\label{eq:14}
\end{equation}

where \(\mathbf{H}_{S}\) and \(\mathbf{H}_{T}\) are real symmetric PEACE Hamiltonians. For each singlet--triplet or distinct-root triplet--triplet pair, an equivariant readout predicts a real Cartesian spin-orbit coupling (SOC) vector with units of energy. Equal-parity pairs use axial-vector \((1e)\ \)features and opposite-parity pairs use polar-vector (\(1o\)) features. Fixed spin-angular coefficients assemble these vectors into complex SOC blocks; reverse blocks are their Hermitian conjugates. Singlet--singlet and same-root triplet SOC blocks are zero in this representation.

We diagonalize \(\mathbf{H}_{S}\) and \(\mathbf{H}_{T}\) separately, obtaining orthogonal eigenvector matrices \(\mathbf{C}_{S}\) and \(\mathbf{C}_{T}\). With \(\mathbf{C}_{sf} = diag\left( \mathbf{C}_{S},\mathbf{C}_{T},\mathbf{C}_{T},\mathbf{C}_{T} \right)\), the SOC matrix in the spin-free adiabatic, or molecular Coulomb Hamiltonian (MCH)\cite{main35}, basis is,

\begin{equation}
\mathbf V^{\mathrm{MCH}}=\mathbf C_{\mathrm{sf}}^\dagger\mathbf V_{\mathrm{SO}}\mathbf C_{\mathrm{sf}},
\label{eq:15}
\end{equation}

\begin{equation}
\mathbf H^{\mathrm{MCH}}=\mathbf E_{\mathrm{sf}}+\mathbf V^{\mathrm{MCH}}.
\label{eq:16}
\end{equation}

Here \(\mathbf{E}_{sf}\) contains the separately diagonalized singlet energies and each triplet energy repeated three times. The predicted SOCs are the entries of \(\mathbf{V}^{MCH}\); the spin-free reference energies and forces are not replaced by eigenvalues or forces of the SOC-mixed Hamiltonian.

\paragraph{Reference data and training.}

To improve sampling of configurations relevant to S\textsubscript{1}--S\textsubscript{0} transitions, the static reference datasets were expanded to 7,500, 7,500 and 15,000 geometries for C\textsubscript{2}H\textsubscript{4}, C\textsubscript{3}H\textsubscript{6}, and C\textsubscript{4}H\textsubscript{8}, using the SA(3)-\allowbreak{}CASSCF(2,2)/\allowbreak{}cc-pVDZ level calculations\cite{main35,main36}. The CH\textsubscript{2}NH\textsubscript{2}\textsuperscript{+} ablation archive contains 4,000 MR-CISD/aug-cc-pVDZ geometries\cite{main37,main38}. Four models were trained separately: full PEACE and variants without the connection, the opposite-parity Hamiltonian block, or both. All four models shared the data split, random seed and initial latent energy--parity ordering as described in \textbf{Supplementary Table 2}. The alkene and CH\textsubscript{2}NH\textsubscript{2}\textsuperscript{+} models used 64 channels for each parity-resolved irreducible representation up to angular momentum \(l_{\max} = 2\), four equivariant message-passing layers, and two equivariant attention layers in each of the Hamiltonian and connection branches. Models were optimized with AdamW\cite{main39} for 1,000 epochs using a batch size of 64. The learning rate followed a cosine schedule from \(10^{- 3}\) to \(10^{- 4}\). For each model, the checkpoint minimizing the validation loss for the physical observables was selected. Energies, forces and smoothed nonadiabatic couplings (SNACs) were fitted using mean-squared-error losses,

\begin{equation}
\mathcal L_E=\left\langle\frac1S\sum_i(\widehat E_i-E_i)^2\right\rangle,
\label{eq:17}
\end{equation}

\begin{equation}
\mathcal L_F=\left\langle\frac1{3NS}\sum_i\left\|\widehat{\mathbf F}_i-\mathbf F_i\right\|_2^2\right\rangle,
\label{eq:18}
\end{equation}

\begin{equation}
\begin{aligned}\mathcal L_{\mathrm{SNAC}}=\Biggl\langle &\min_{\substack{s_i\in\{-1,+1\}\\s_0=1}}\frac1{3Nn_p}\\[-1pt] &{}\times\sum_{i<j}\left\|\widehat{\mathbf b}_{ij}-s_is_j\mathbf b_{ij}\right\|_2^2\Biggr\rangle,\end{aligned}
\label{eq:19}
\end{equation}

where \(S\) is the number of electronic states, \(N\) is the number of atoms, \(n_{p} = S(S - 1)/2\), and angular brackets denote averaging over geometries. Hats indicate predictions. The SNAC loss minimizes over a single consistent set of electronic-state signs for all state pairs within each geometry\cite{main20}. An additional energy-gap loss was included,

\begin{equation}
\begin{aligned}\mathcal L_\Delta=\Biggl\langle\frac1{n_p}\sum_{i<j}\Bigl[&(\widehat E_j-\widehat E_i)\\[-1pt] &{}-(E_j-E_i)\Bigr]^2\Biggr\rangle.\end{aligned}
\label{eq:20}
\end{equation}

The total objective was

\begin{equation}
\begin{aligned}\mathcal L={}&\lambda_E\mathcal L_E+\lambda_F\mathcal L_F+\lambda_{\mathrm{SNAC}}\mathcal L_{\mathrm{SNAC}}\\ &+\lambda_\Delta\mathcal L_\Delta+\mathcal L_{\mathrm{parity}},\end{aligned}
\label{eq:21}
\end{equation}

with all four observable weights set numerically to 1 using energies in eV and forces and SNACs in eV Å\textsuperscript{-1}. For reflection-symmetric training configurations, including reflections combined with permutations of identical atoms, relative electronic-state parities were inferred from the symmetry selection rules of the reference SNACs. Configurations with confidently assigned labels received auxiliary supervision of the parity-sector spectra and signed energy gaps between opposite-parity states. The corresponding losses per geometry were,

\begin{equation}
\ell_{\mathrm{sec}}^{(q)}=\left\langle(\widehat E_a^{(q)}-E_a^{(q)})^2\right\rangle_a,
\label{eq:22}
\end{equation}

\begin{equation}
\ell_{\mathrm{rel}}^{(q)}=\left\langle\frac{(\widehat\Delta_{ab}^{(q)}-\Delta_{ab}^{(q)})^2}{(\Delta_{ab}^{(q)})^2+\epsilon^2}\right\rangle_{a+,b-}.
\label{eq:23}
\end{equation}

Here, \({\widehat{E}}_{a}^{(q)}\) denotes a sorted eigenvalue of a parity block matched to a labelled reference level under assignment \(q\), and \(\Delta_{ab}^{(q)} = E_{b}^{(q)} - E_{a}^{(q)}\). Only labelled levels and labelled opposite-parity pairs contribute. Because the labels determine relative parity, the auxiliary objective was minimized over the two globally reversed assignments,

\begin{equation}
\begin{aligned}
\mathcal{L}_{\mathrm{parity}}
  &= \Biggl\langle \min_{q\in\{+,-\}}\Bigl[
       \lambda_{\mathrm{sec}}\ell_{\mathrm{sec}}^{(q)}\\
  &\quad{}+\lambda_{\mathrm{rel}}\ell_{\mathrm{rel}}^{(q)}
       +\lambda_{\mathrm{ord}}\ell_{\mathrm{ord}}^{(q)}
       \Bigr]\Biggr\rangle_{\mathrm{labelled}} .
\end{aligned}
\label{eq:24}
\end{equation}

The ordering term averages squared positive violations of the reference ordering between opposite-sector levels. Unlabelled latent levels are assigned ranks above the observed window.

\paragraph{Nonadiabatic molecular dynamics.}

Surface-hopping simulations\cite{main40} were performed using the SHARC 4 framework\cite{main41,main42}. For CH\textsubscript{2}NH\textsubscript{2}\textsuperscript{+}, the reference ensemble comprised 145 MR-CISD/aug-cc-pVDZ trajectories\cite{main37,main38} with a maximum absolute total-energy drift below 0.25 eV. The PEACE, no-connection, no-odd and double-ablation models were evaluated using the same 1,000 Wigner-sampled initial conditions, with trajectories initiated in S\textsubscript{2} and propagated for 100 fs. Only complete trajectories with a maximum absolute total-energy drift below 0.25 eV were retained. For alkenes, the reference ensembles each comprised 200 trajectories with a maximum absolute total-energy drift below 0.25 eV calculated at the SA (3)-CASSCF (2,2)/cc-pVDZ level\cite{main35,main36}. For each alkene, 1,000 PEACE trajectories were initiated in S\textsubscript{1} from Wigner-sampled initial conditions and propagated for 100 fs using the same propagation settings and trajectory-selection criteria. For CH\textsubscript{2}S, 1,000 PEACE trajectories were initiated in S\textsubscript{1} from Wigner-sampled initial conditions and propagated for 3 ps, with both nonadiabatic and spin--orbit couplings included. All calculations use a nuclear time step of 0.5 fs and 25 electronic substeps per nuclear step. The state populations were calculated from the retained ensembles, with pointwise 95\% confidence intervals estimated by trajectory-level bootstrap resampling. Trajectory selection was performed independently for the reference and model ensembles.

\section*{Code availability}

The implementation of PEACE code and trained models are available via GitHub at \url{https://github.com/solvendus/peace}.

\section*{Acknowledgement}

We gratefully acknowledge the financial support from Hong Kong Quantum AI Lab Limited, Air @ InnoHK of Hong Kong Government, GuangDong Basic and Applied Basic Research Foundation (Grant No. 2024A1515013283), and the National Natural Science Foundation of China (Grant No. 22473022).

\ifincludesi
\clearpage
\nobalance
\onecolumn
\setcounter{figure}{0}
\setcounter{table}{0}
\renewcommand{\theHfigure}{supp.\arabic{figure}}
\renewcommand{\theHtable}{supp.\arabic{table}}
\captionsetup[figure]{name=Supplementary Fig.}
\renewcommand{\figurename}{Supplementary Fig.}
\captionsetup[table]{name=Supplementary Table}
\renewcommand{\tablename}{Supplementary Table}
\captionsetup{skip=4pt,hypcap=false}

\SIpageblock{intro and S1}{%
\begin{center}
{\Large\bfseries Supplementary Information\par}
\vspace{6pt}
{\large\bfseries PEACE: Covariant learning of nonadiabatic manifolds\\
with parity-resolved Hamiltonians\par}
\end{center}
\begin{multicols}{2}
\section*{Reference data and comparison models}

The alkene reference calculations follow the electronic-structure level used in the SPaiNN benchmark,\cite{si1} with three singlet states at SA(3)-\allowbreak{}CASSCF(2,2)/\allowbreak{}cc-pVDZ.\cite{si2,si3} Final expanded datasets contain 7,500, 7,500 and 15,000 geometries for C\textsubscript{2}H\textsubscript{4}, C\textsubscript{3}H\textsubscript{6}, and C\textsubscript{4}H\textsubscript{8}, respectively. The CH\textsubscript{2}NH\textsubscript{2}\textsuperscript{+} ablation dataset contains 4,000 MR-CISD geometries, with energies, gradients and derivative couplings evaluated using the aug-cc-pVDZ basis.\cite{si4,si5,si6} Training, validation and test counts are given in \textbf{Supplementary Table 2}.

The external set of 770 CH\textsubscript{2}NH\textsubscript{2}\textsuperscript{+} geometries in \textbf{Supplementary Fig. 5} originates from the SchNarc benchmark.\cite{si7} All compared checkpoints were evaluated on the same geometries. For CH\textsubscript{2}S, the reference labels were recomputed with OpenMolcas\cite{si8} at the CASSCF(6,5)/def2-SVP level\cite{si2,si9}. Spin--orbit couplings were obtained with atomic mean-field spin--orbit integrals\cite{si10,si11}. The 4,000/200/503 split follows the counts used in the SchNarc benchmark\cite{si7} with independently reconstructed membership.

The SPaiNN comparisons use SchNet and PaiNN architectures\cite{si1,si12,si13}. Alkene and CH\textsubscript{2}NH\textsubscript{2}\textsuperscript{+} comparison models were retrained on the datasets with the same batch size, splits and number of epochs as PEACE. Exciting DeePMD uses the public checkpoint with coupling-dyad representation\cite{si14}.

\section*{Timing and numerical symmetry tests}

The timing benchmark uses 4,000 configurations from 20 complete reference trajectories per alkene. PEACE evaluations were repeated three times at batch size one after model loading and JIT warm-up. Timings include energy, force and nonadiabatic coupling evaluation, device synchronization, transfer to the host and the native SHARC interface work. Archived CASSCF timings retain their native electronic-structure and interface workloads. The reported ratios therefore apply to the specified CPU/GPU configurations. Timing intervals use 10,000 paired bootstrap resamples of whole trajectory blocks\cite{si16}. FP64 denotes inference precision; the checkpoints were trained in FP32 and promoted to FP64 for this benchmark. \textbf{Supplementary Fig. 15} tests numerical transformation consistency under the specified symmetry operations; predictive accuracy is assessed separately by the reference comparisons.
\end{multicols}
\vspace{5pt}
\SIfigure{width=0.86\textwidth}{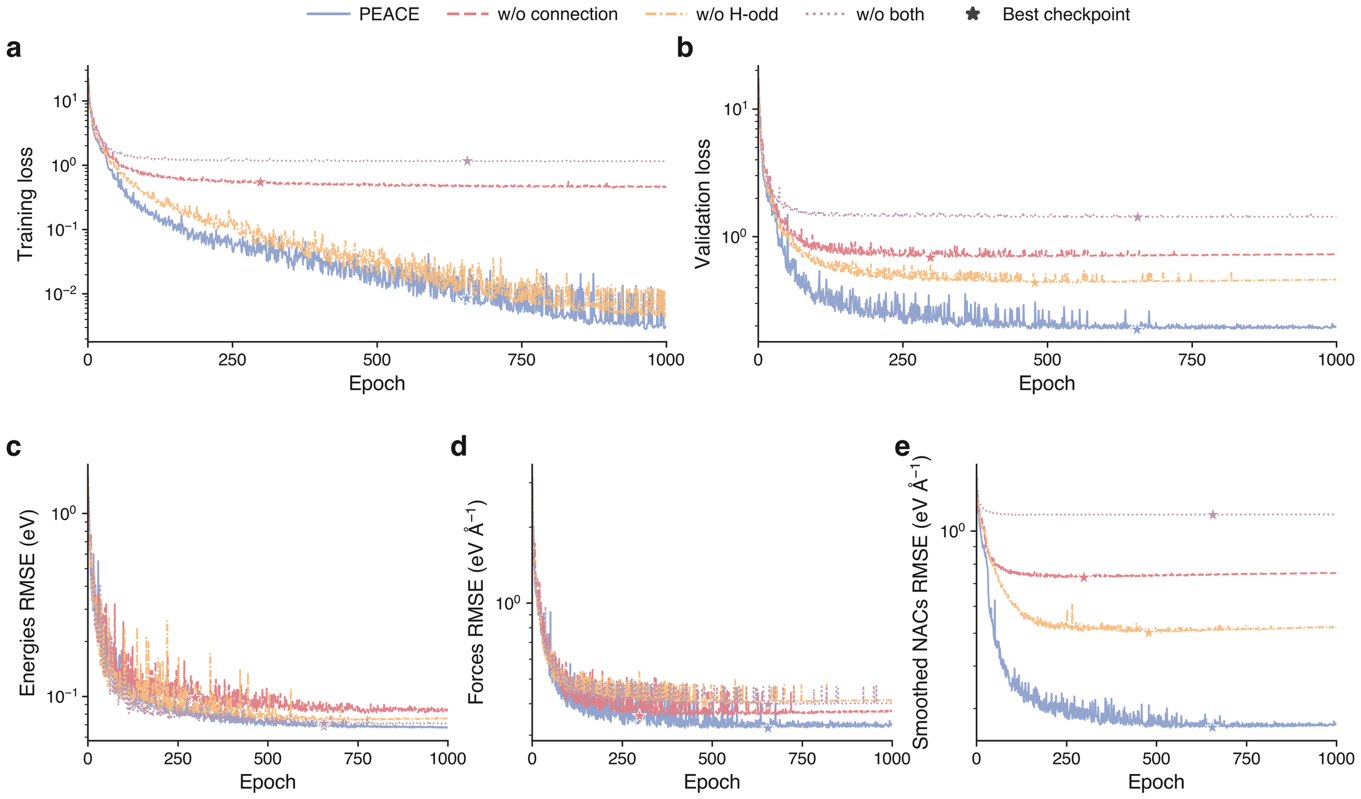}{\textbf{Learning curves for PEACE and the ablation models.} \textbf{a}, Training loss. \textbf{b}, Validation loss used for checkpoint selection. \textbf{c}--\textbf{e}, Validation root-mean-square errors (RMSEs) for energies, Cartesian force components and smoothed nonadiabatic coupling (SNAC) components, respectively. PEACE is compared with models excluding the connection contribution (w/o connection), opposite-parity Hamiltonian blocks (w/o H-odd), or both (w/o both). Each model was trained for 1,000 epochs using 2,500 training and 250 validation geometries. Stars mark the validation-selected checkpoints.}{suppfig:1}
}

\newpage

\SIpageblock{S2}{%
\SIfigure{width=0.86\textwidth}{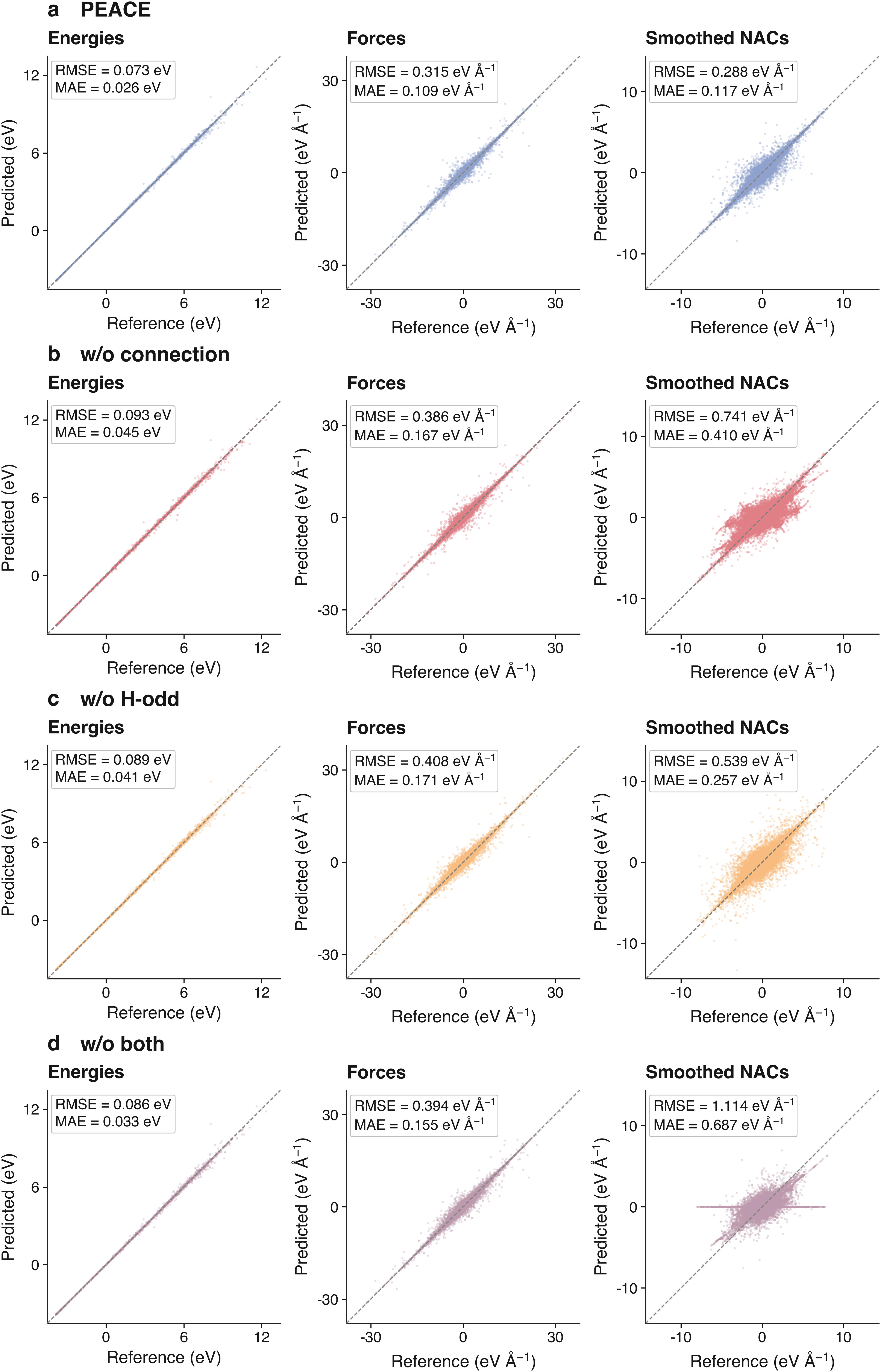}{\textbf{Static test predictions for the CH\textsubscript{2}NH\textsubscript{2}\textsuperscript{+} ablations.} Rows \textbf{a}--\textbf{d} correspond to PEACE, w/o connection, w/o H-odd and w/o both, respectively. Columns show predicted versus MR-CISD/aug-cc-pVDZ reference energies, Cartesian force components and SNAC components\cite{si4,si5,si6}. All four models use validation-selected checkpoints and the same 1,250 held-out test geometries. Insets report mean absolute errors (MAE) and root mean squared errors (RMSE).}{suppfig:2}
}

\newpage

\SIpageblock{S3 to S5}{%
\SIfigure{width=0.77\textwidth}{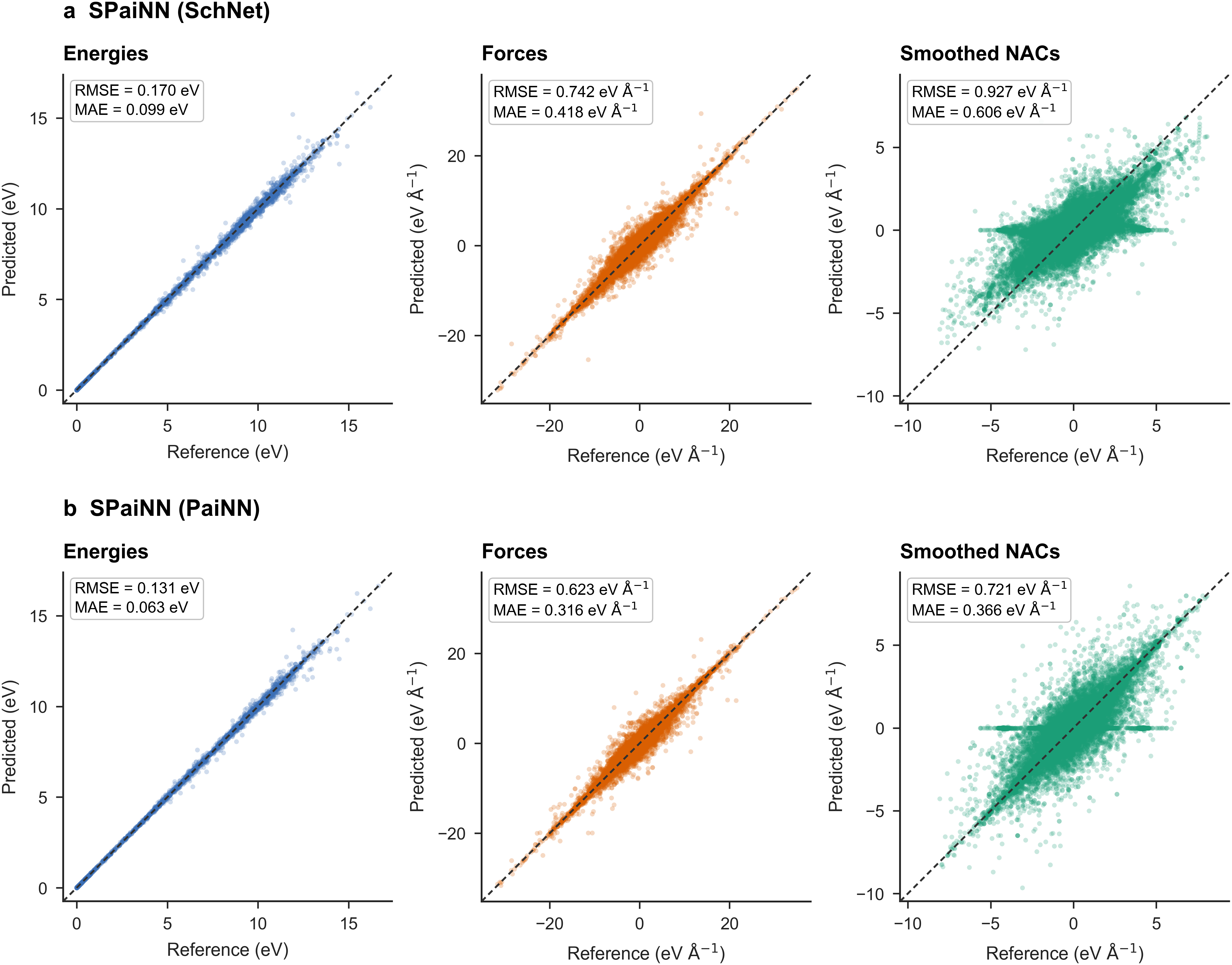}{\textbf{Static test predictions of SPaiNN models for CH\textsubscript{2}NH\textsubscript{2}\textsuperscript{+}.} SPaiNN\cite{si1} models use \textbf{a}, SchNet\cite{si12} and \textbf{b}, PaiNN\cite{si13} architectures. Columns show predicted versus MR-CISD/aug-cc-pVDZ reference energies, Cartesian force components and SNAC components. Both models use validation-selected checkpoints and the same 1,250 held-out test geometries as \textbf{Supplementary Fig. 2}. Insets report MAE and RMSE.}{suppfig:3}
\par\vspace{5pt}
\SIfigure{width=\textwidth}{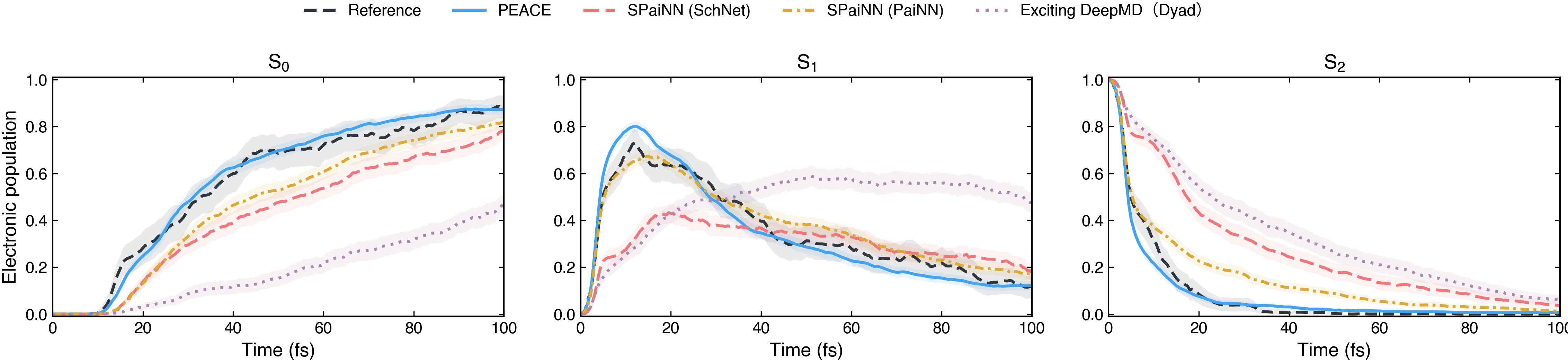}{\textbf{CH\textsubscript{2}NH\textsubscript{2}\textsuperscript{+} population dynamics.} Reference S\textsubscript{0}, S\textsubscript{1} and S\textsubscript{2} populations are compared with PEACE, SPaiNN--SchNet, SPaiNN--PaiNN\cite{si1} and Exciting DeePMD Dyad\cite{si14} over 100 fs. The four machine-learning methods use the same 1,000 initial geometries and velocities, with trajectories initiated in S\textsubscript{1}. Independent selection for complete trajectories with a maximum absolute total-energy drift below 0.25 eV retains 947, 418, 758 and 453 model trajectories, respectively. Shading indicates pointwise 95\% confidence intervals from trajectory-level bootstrap resampling\cite{si16}.}{suppfig:4}
\par\vspace{5pt}
\SIfigure{width=\textwidth}{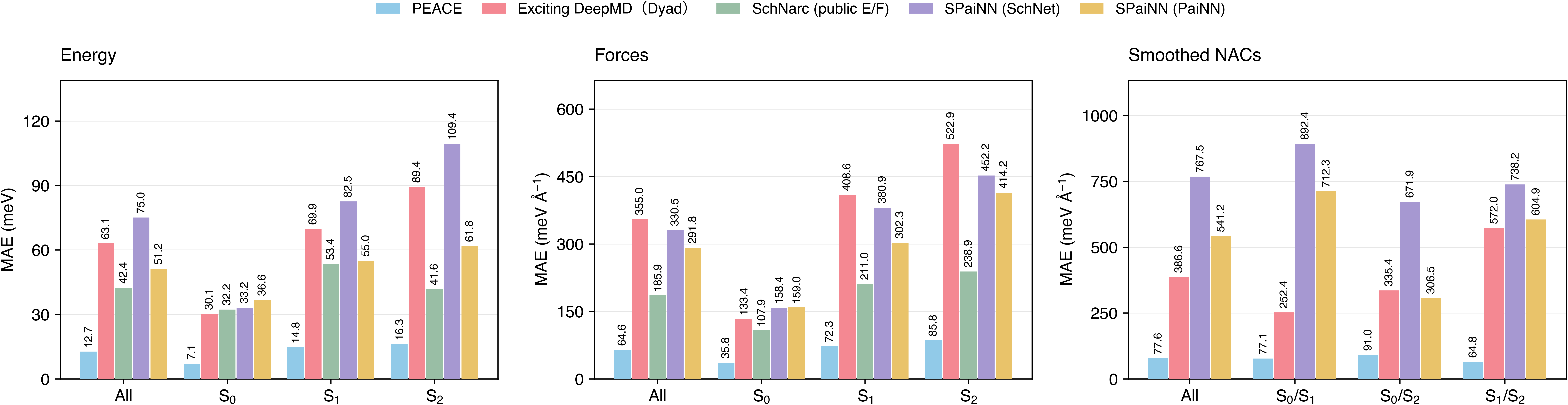}{\textbf{Unseen 770-geometry CH\textsubscript{2}NH\textsubscript{2}\textsuperscript{+} benchmark.} Energy, force and SNAC MAE are evaluated on the same 770 geometries from the SchNarc benchmark\cite{si7}. Comparisons include PEACE, Exciting DeePMD\cite{si14}, SchNarc\cite{si7} and SchNet- and PaiNN-based SPaiNN models\cite{si1}. Force and SNAC errors are averaged over Cartesian components. Each predicted SNAC vector was sign-aligned independently for each geometry and state pair by choosing the sign that minimized its error relative to the reference vector. The public SchNarc\cite{si7} checkpoint has no NAC prediction head, so no SNAC error is reported.}{suppfig:5}
}

\newpage

\SIpageblock{S6 and S7}{%
\SIfigure{width=0.71\textwidth}{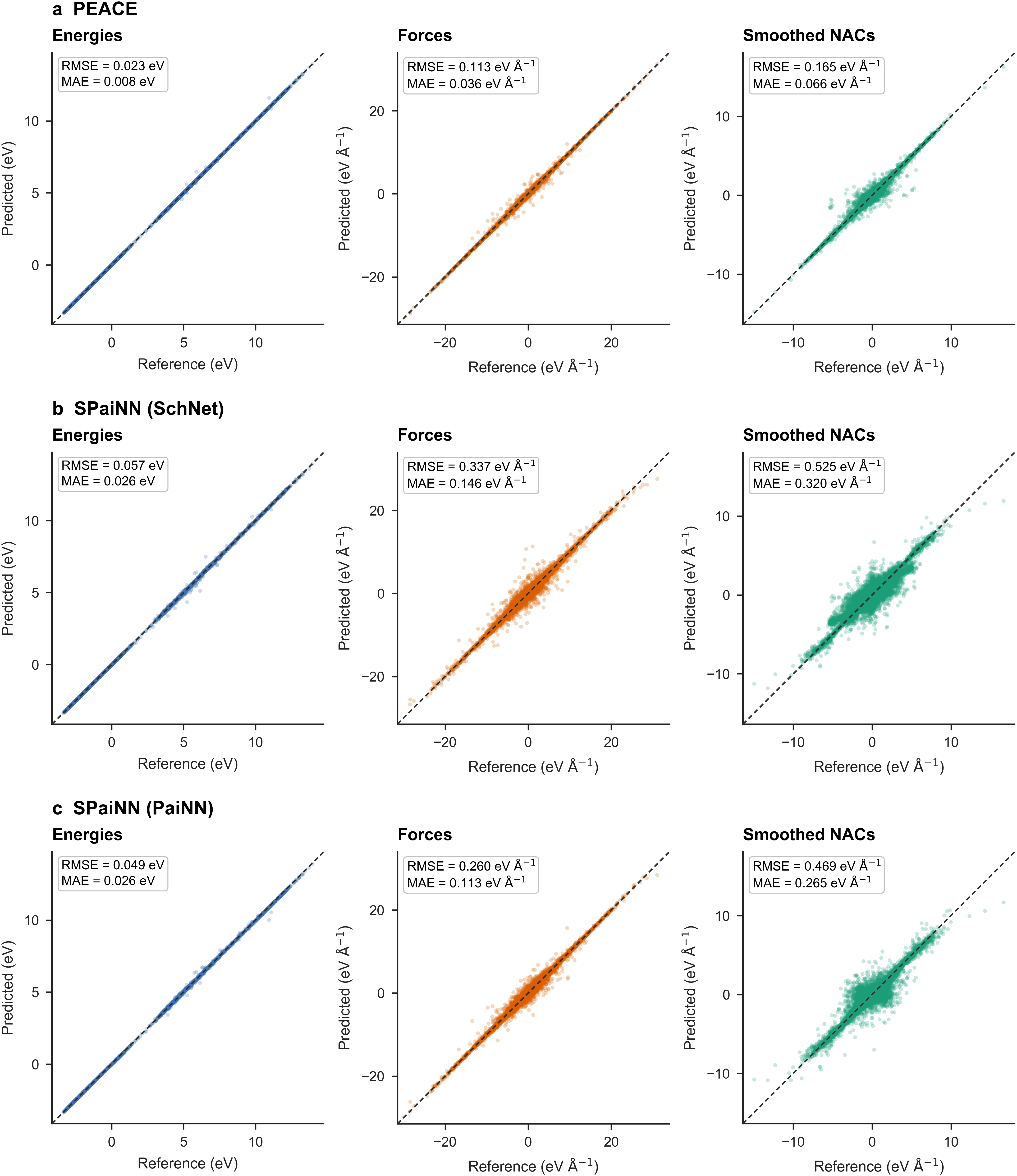}{\textbf{Static predictions for C\textsubscript{2}H\textsubscript{4}.} Predicted versus SA(3)-\allowbreak{}CASSCF(2,2)/\allowbreak{}cc-pVDZ reference energies, Cartesian force components and SNAC components for \textbf{a}, PEACE; \textbf{b}, SPaiNN--SchNet; and \textbf{c}, SPaiNN--PaiNN.\cite{si1} Dashed lines indicate equality. Insets report MAE and RMSE for the test set.}{suppfig:6}
\par\vspace{6pt}
\SIsidefigure{width=\linewidth}{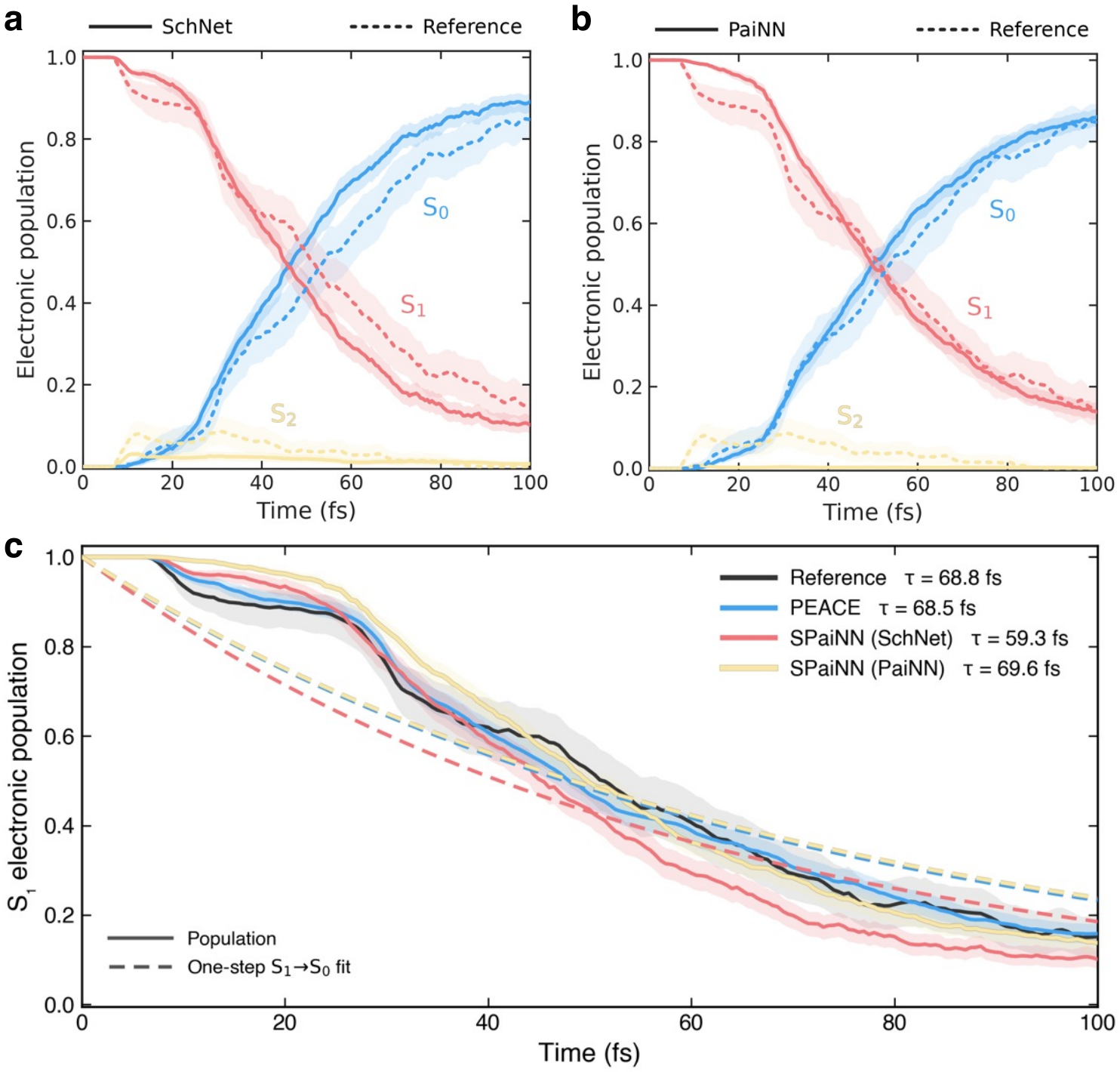}{\textbf{Electronic-state populations and S\textsubscript{1} relaxation dynamics of C\textsubscript{2}H\textsubscript{4}.} \textbf{a}, \textbf{b}, SPaiNN populations using SchNet (\textbf{a}) and PaiNN (\textbf{b})\cite{si1}. Solid lines show model predictions and dotted lines the SA(3)-\allowbreak{}CASSCF(2,2)/\allowbreak{}cc-pVDZ reference. \textbf{c}, S\textsubscript{1} populations from the reference (black), PEACE (blue), SPaiNN--SchNet (red) and SPaiNN--PaiNN (yellow). Solid lines show ensemble means; dashed lines show fits to the effective one-step \(S_{1} \rightarrow S_{0}\) model, \(P_{S_{1}}(t) = \exp( - t/\tau)\), over 0--100 fs. Fitted decay times are 68.8, 68.5, 59.3 and 69.6 fs, respectively. Shading indicates pointwise 95\% confidence intervals from trajectory-level bootstrap resampling\cite{si16}. Trajectories start in S\textsubscript{1}; only complete trajectories with a maximum absolute total-energy drift below 0.25 eV are retained.}{suppfig:7}
}

\newpage

\SIpageblock{S8 and S9}{%
\SIfigure{width=0.71\textwidth}{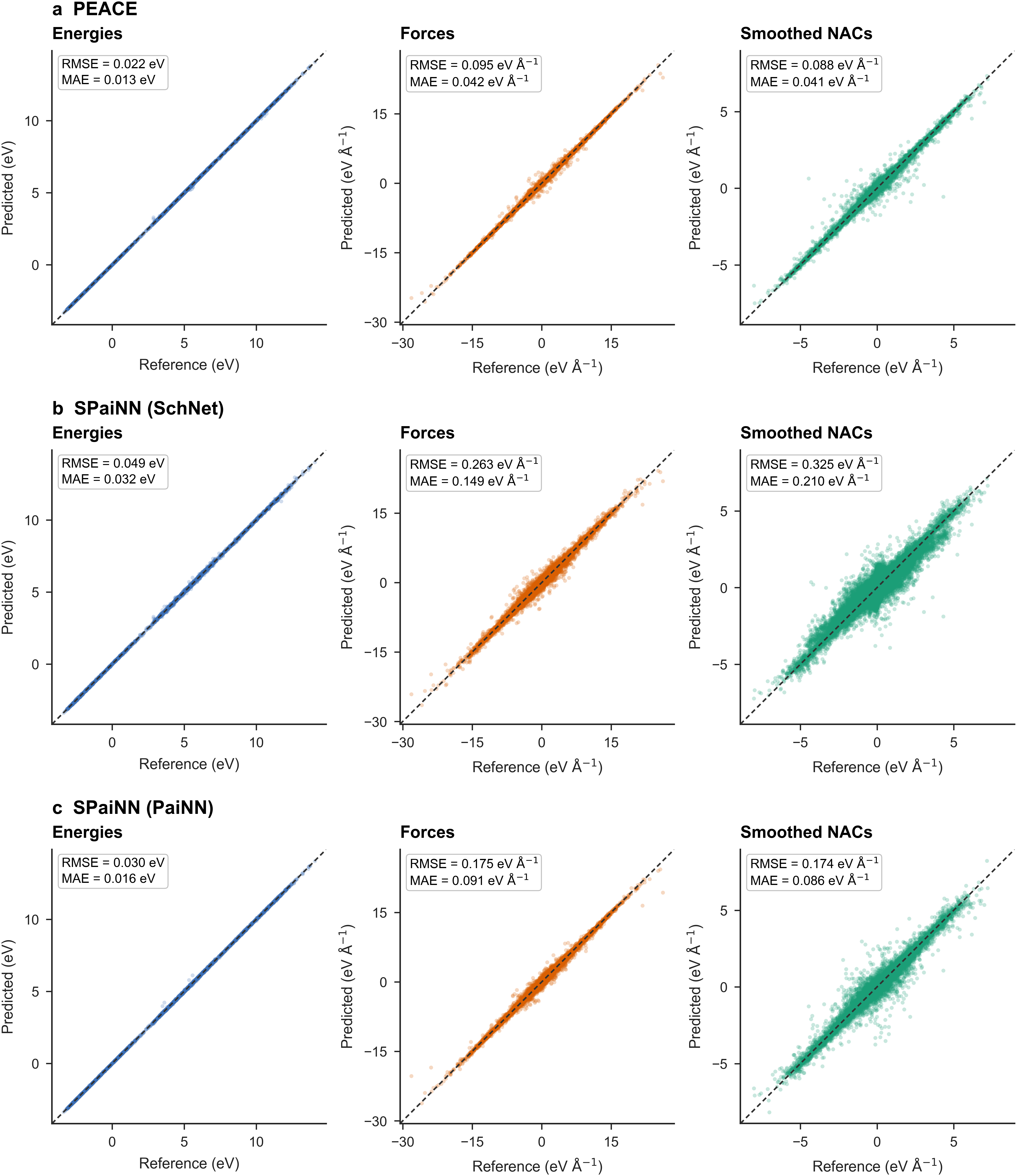}{\textbf{Static predictions for C\textsubscript{3}H\textsubscript{6}.} Predicted versus SA(3)-\allowbreak{}CASSCF(2,2)/\allowbreak{}cc-pVDZ reference energies, Cartesian force components and SNAC components for \textbf{a}, PEACE; \textbf{b}, SPaiNN--SchNet; and \textbf{c}, SPaiNN--PaiNN.\cite{si1} Dashed lines indicate equality. Insets report MAE and RMSE for the test set.}{suppfig:8}
\par\vspace{6pt}
\SIsidefigure{width=\linewidth}{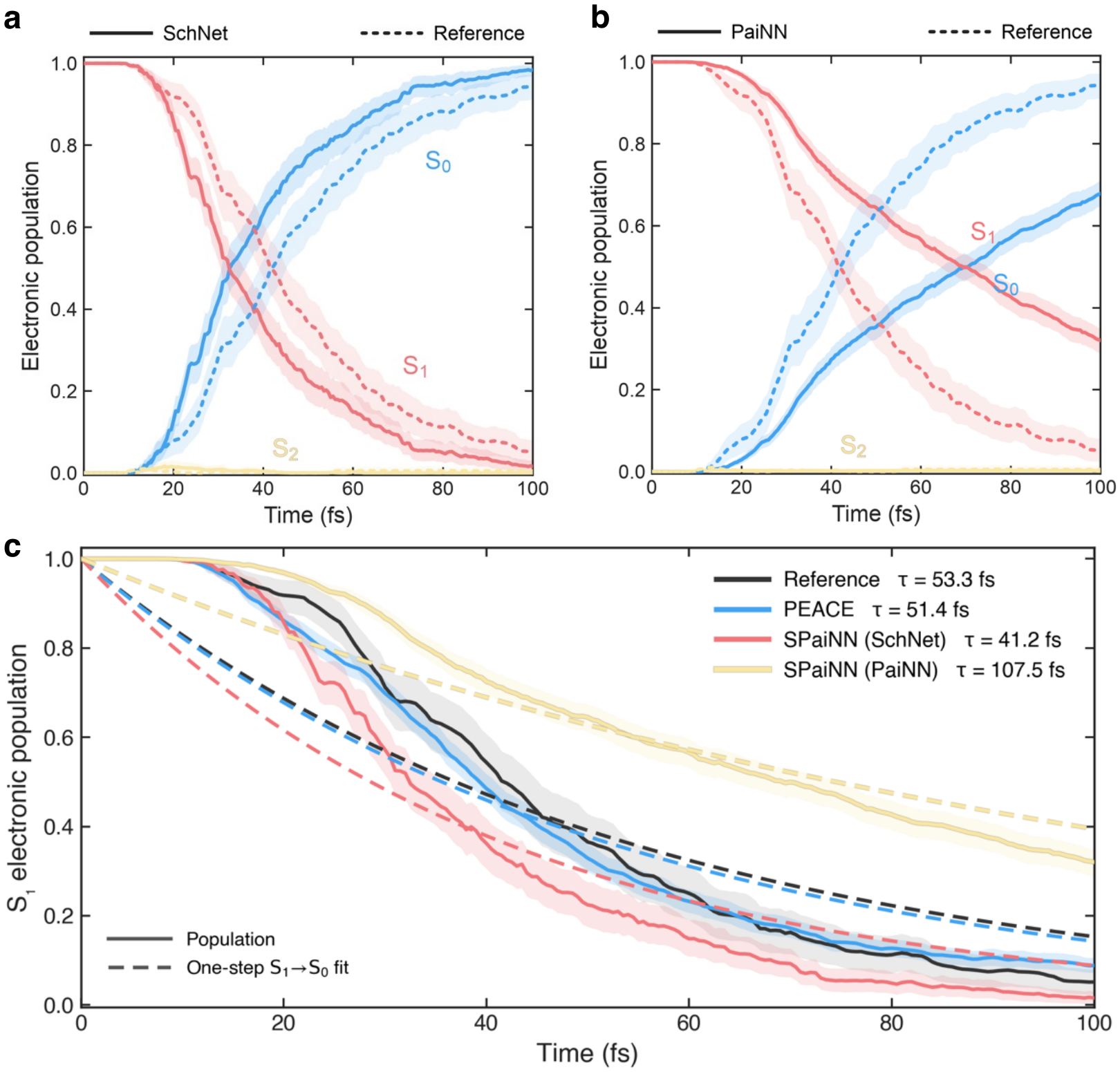}{\textbf{Electronic-state populations and S\textsubscript{1} relaxation dynamics of C\textsubscript{3}H\textsubscript{6}.} \textbf{a}, \textbf{b}, SPaiNN populations using SchNet (\textbf{a}) and PaiNN (\textbf{b})\cite{si1}. Solid lines show model predictions and dotted lines the SA(3)-\allowbreak{}CASSCF(2,2)/\allowbreak{}cc-pVDZ reference. \textbf{c}, S\textsubscript{1} populations from the reference (black), PEACE (blue), SPaiNN--SchNet (red) and SPaiNN--PaiNN (yellow). Solid lines show ensemble means; dashed lines show fits to the effective one-step \(S_{1} \rightarrow S_{0}\) model, \(P_{S_{1}}(t) = \exp( - t/\tau)\), over 0--100 fs. Fitted decay times are 53.3, 51.4, 41.2 and 107.5 fs, respectively. Shading indicates pointwise 95\% confidence intervals from trajectory-level bootstrap resampling\cite{si16}. Trajectories start in S\textsubscript{1}; only complete trajectories with a maximum absolute total-energy drift below 0.25 eV are retained.}{suppfig:9}
}

\newpage

\SIpageblock{S10 and S11}{%
\SIfigure{width=0.71\textwidth}{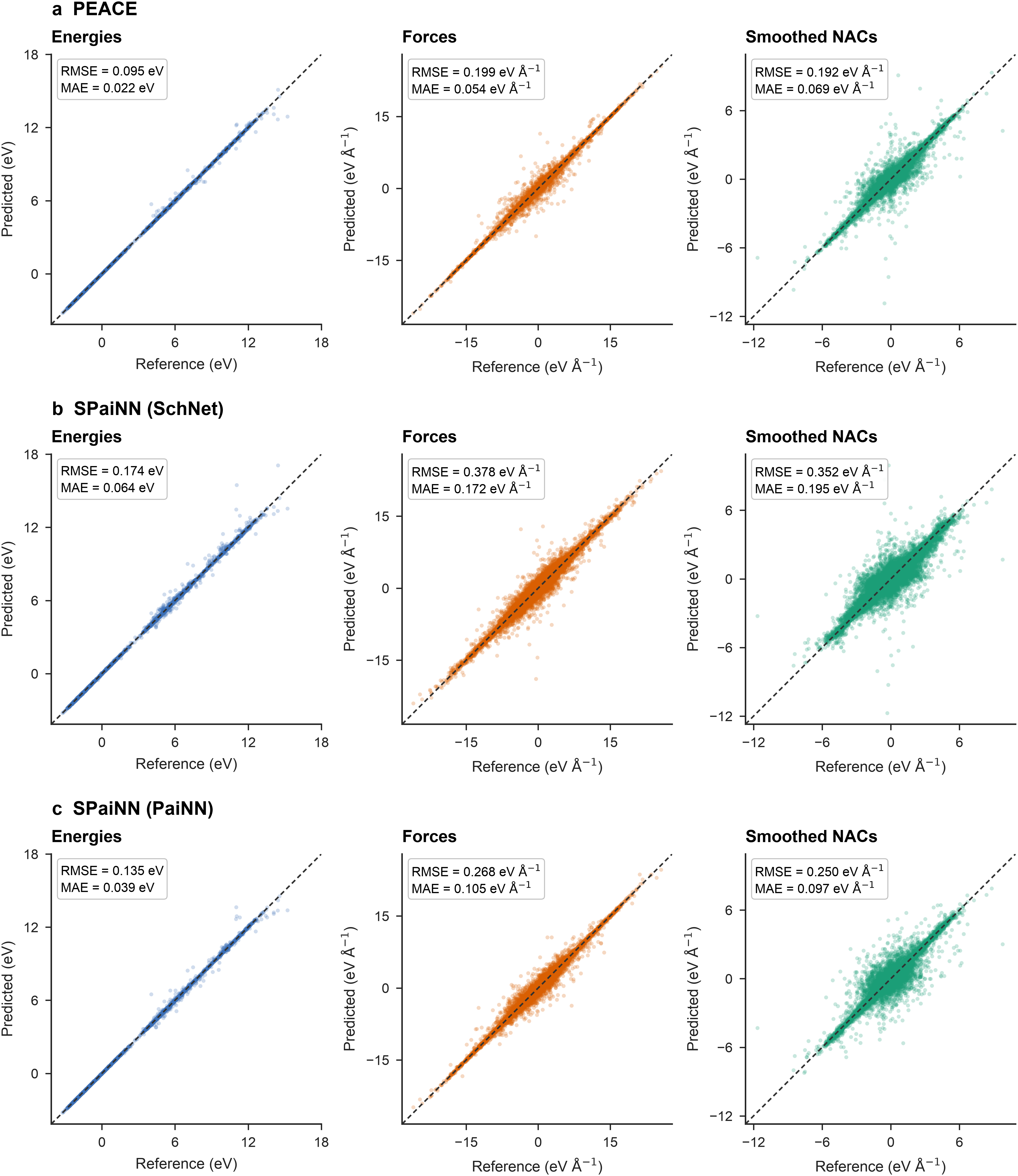}{\textbf{Static predictions for C\textsubscript{4}H\textsubscript{8}.} Predicted versus SA(3)-\allowbreak{}CASSCF(2,2)/\allowbreak{}cc-pVDZ reference energies, Cartesian force components and SNAC components for \textbf{a}, PEACE; \textbf{b}, SPaiNN--SchNet; and \textbf{c}, SPaiNN--PaiNN\cite{si1}. Dashed lines indicate equality. Insets report MAE and RMSE for the test set.}{suppfig:10}
\par\vspace{6pt}
\SIsidefigure{width=\linewidth}{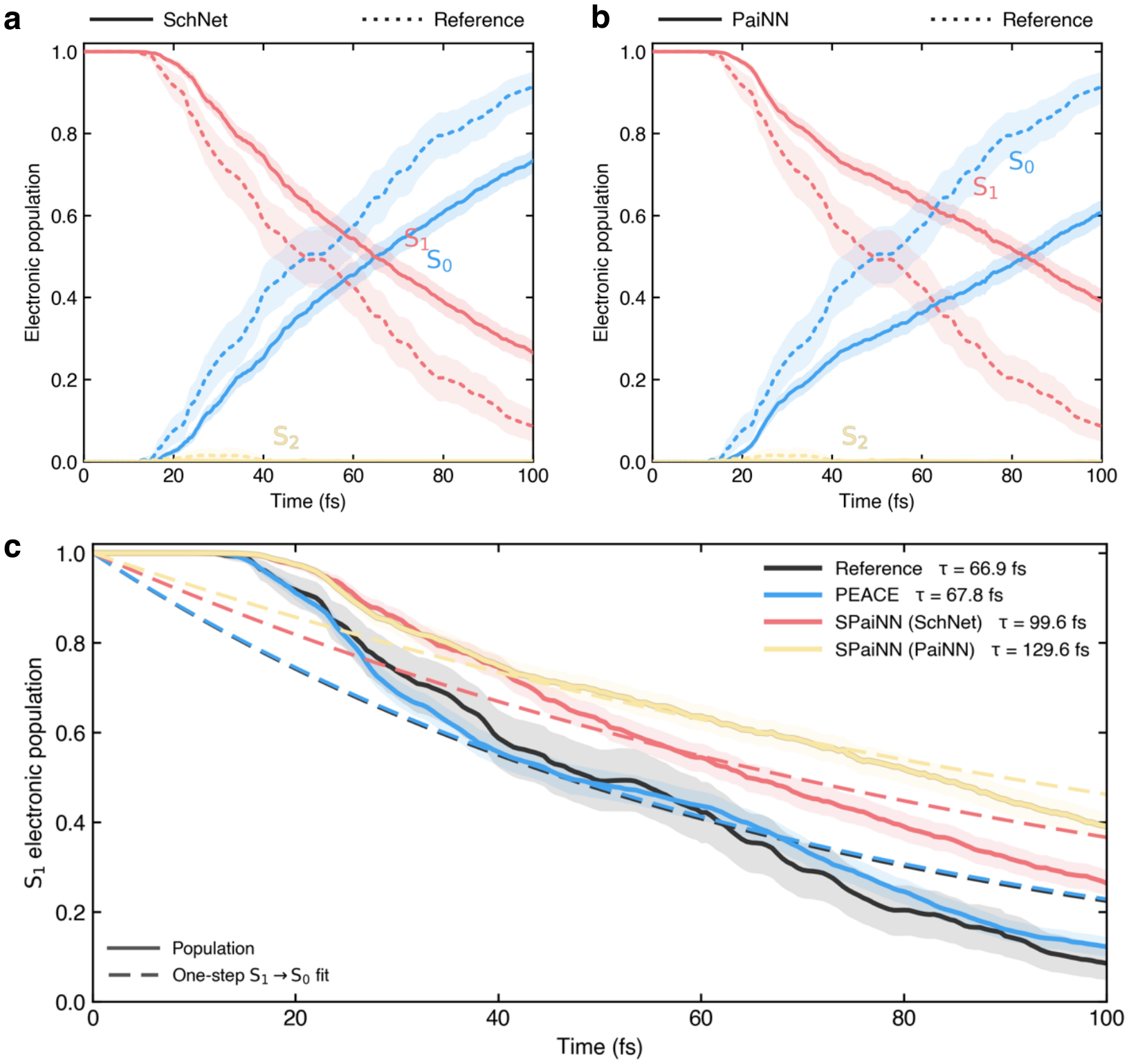}{\textbf{Electronic-state populations and S\textsubscript{1} relaxation dynamics of C\textsubscript{4}H\textsubscript{8}.} \textbf{a}, \textbf{b}, SPaiNN populations using SchNet (\textbf{a}) and PaiNN (\textbf{b})\cite{si1}. Solid lines show model predictions and dotted lines the SA(3)-\allowbreak{}CASSCF(2,2)/\allowbreak{}cc-pVDZ reference. \textbf{c}, S\textsubscript{1} populations from the reference (black), PEACE (blue), SPaiNN--SchNet (red) and SPaiNN--PaiNN (yellow). Solid lines show ensemble means; dashed lines show fits to the effective one-step \(S_{1} \rightarrow S_{0}\) model, \(P_{S_{1}}(t) = \exp( - t/\tau)\), over 0--100 fs. Fitted decay times are 66.9, 67.8, 99.6 and 129.6 fs, respectively. Shading indicates pointwise 95\% confidence intervals from trajectory-level bootstrap resampling\cite{si16}. Trajectories start in S\textsubscript{1}; only complete trajectories with a maximum absolute total-energy drift below 0.25 eV are retained.}{suppfig:11}
}

\newpage

\SIpageblock{S12 and S13}{%
\SIfigure{width=0.85\textwidth}{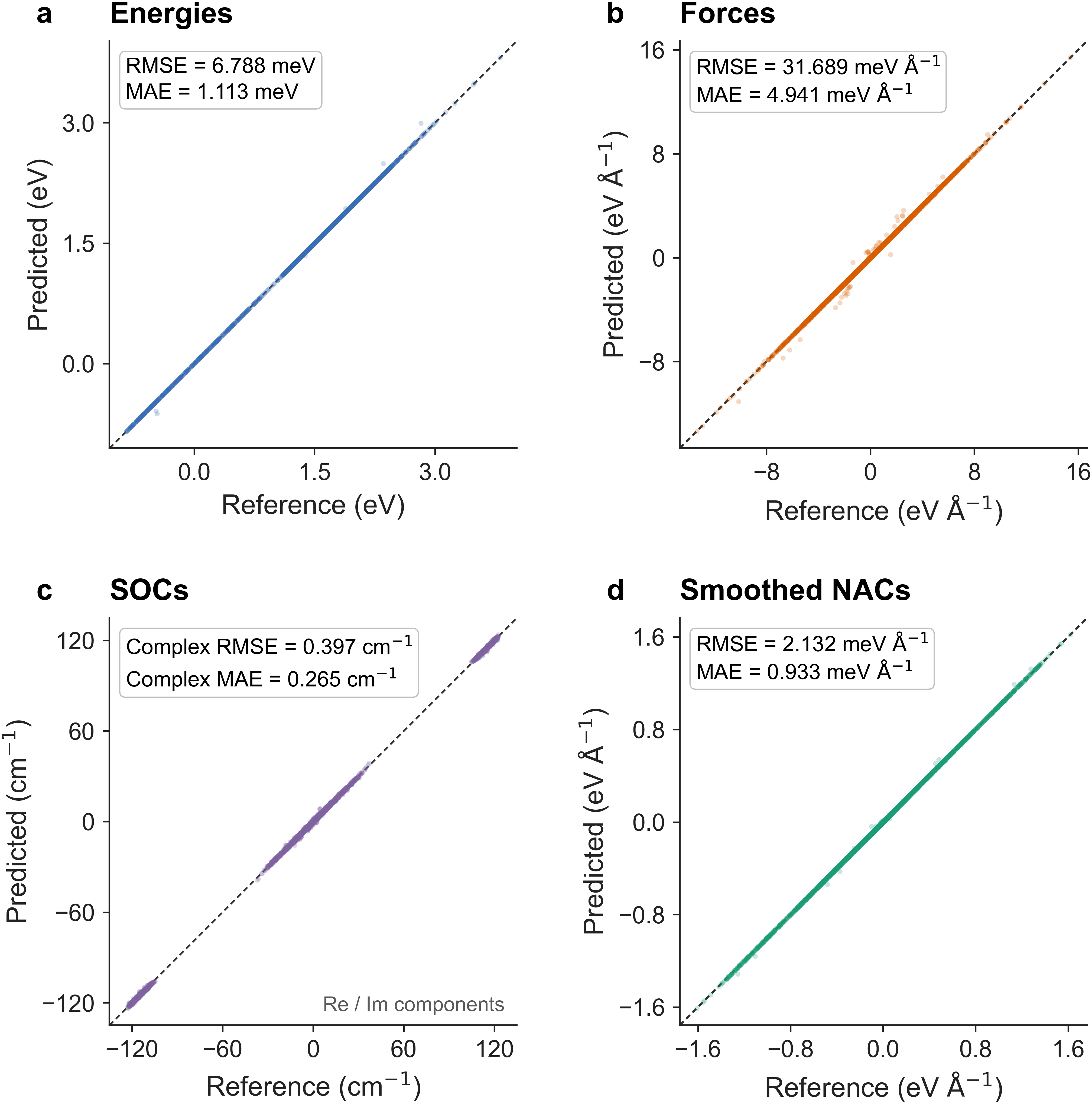}{\textbf{Static predictions for CH\textsubscript{2}S including spin--orbit coupling.} Predicted versus independently recomputed OpenMolcas reference values\cite{si8} for \textbf{a}, energies; \textbf{b}, Cartesian force components; \textbf{c}, complex spin--orbit couplings (SOCs); and \textbf{d}, SNAC components on the test set. In \textbf{c}, real and imaginary parts are plotted separately, whereas the inset errors use the modulus of the complex difference.}{suppfig:12}
\par\vspace{6pt}
\SIfigure{width=\textwidth}{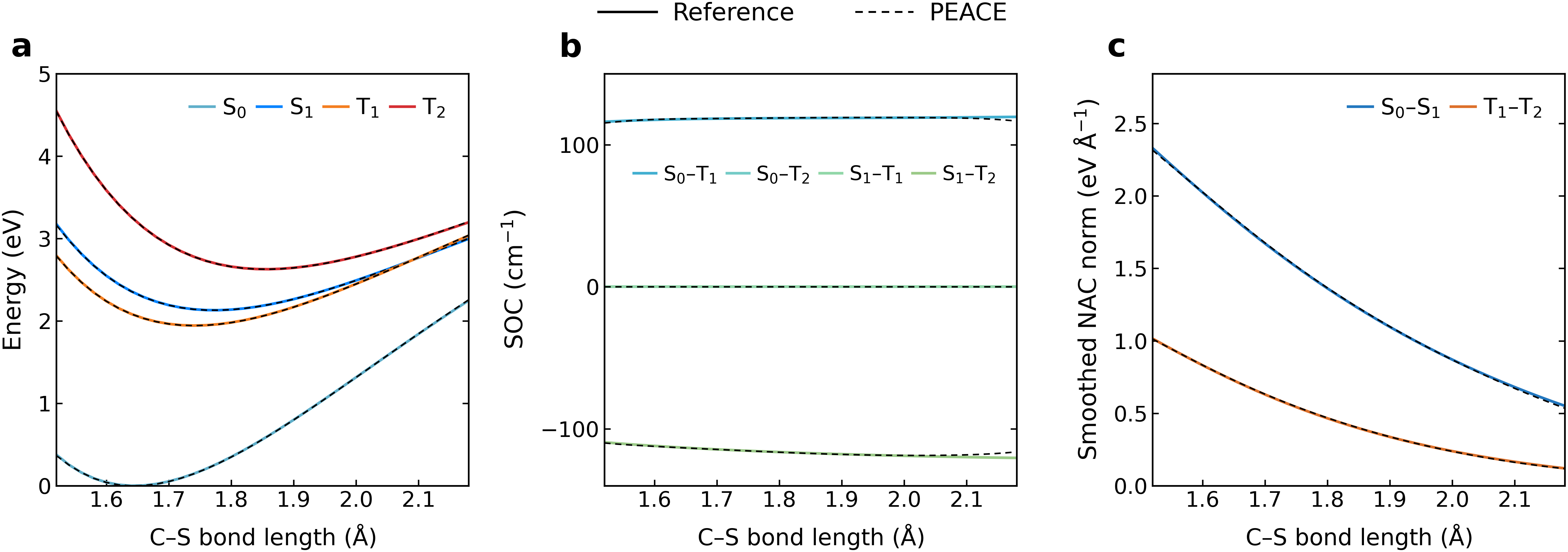}{\textbf{CH\textsubscript{2}S properties along a C--S bond-length scan.} PEACE predictions and independent CASSCF(6,5)/def2-SVP OpenMolcas reference calculations\cite{si8} are compared for \textbf{a}, electronic energies; \textbf{b}, selected signed SOC components; and \textbf{c}, SNAC norms. Panel \textbf{b} shows Re(SOC) + Im(SOC) for the \emph{M}\textsubscript{s} = +1 triplet component. Panel \textbf{c} shows Euclidean norms over all 12 Cartesian components. The 34 plotted geometries span C--S distances of 1.52--2.18 Å.}{suppfig:13}
}

\newpage

\SIpageblock{S14 S15 and tables}{%
\SIhalffigure{width=\linewidth}{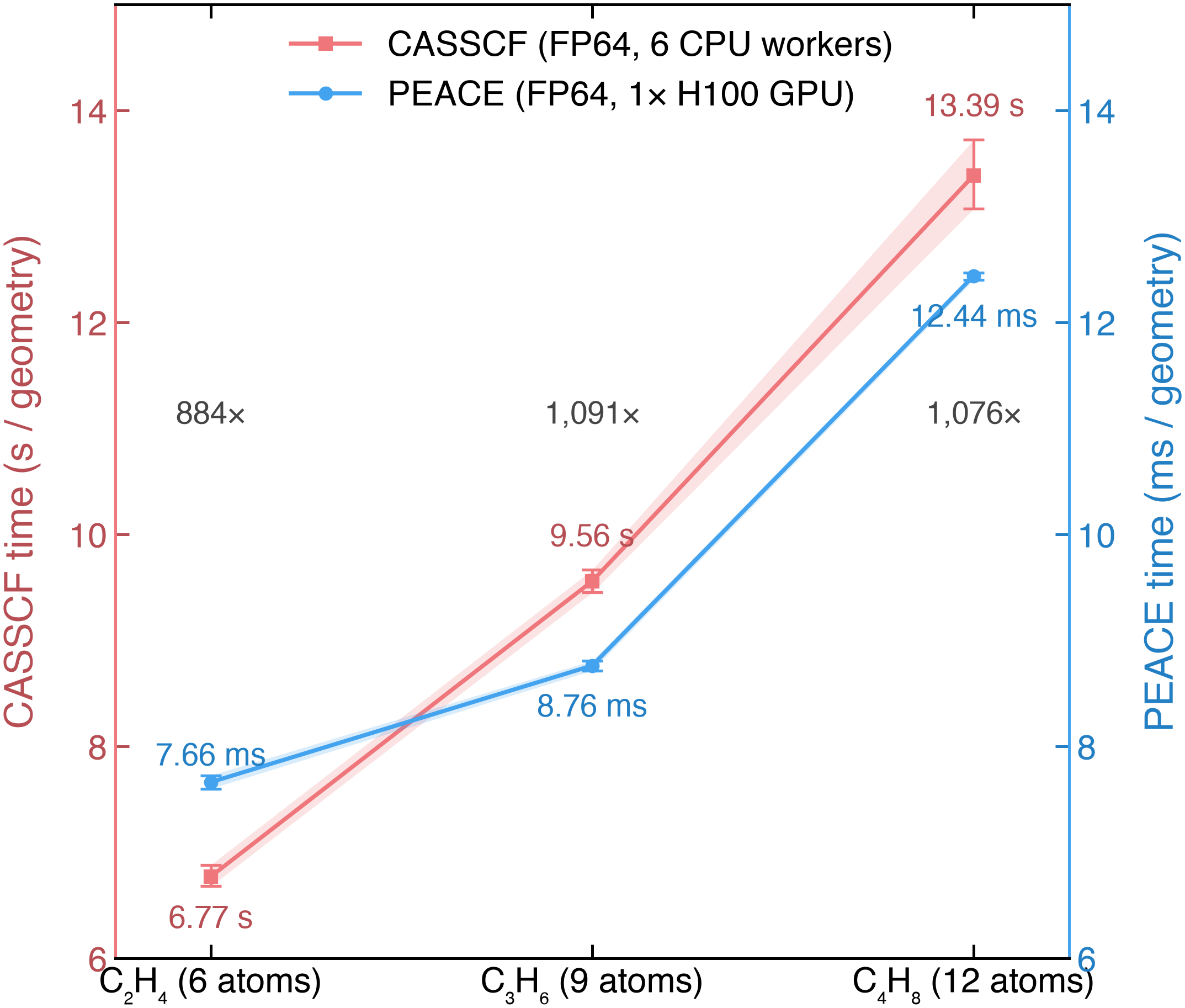}{\textbf{Per-geometry electronic-evaluation times for PEACE and CASSCF.} FP64 PEACE inference at batch size one on a single NVIDIA H100 GPU is compared with SA(3)-\allowbreak{}CASSCF(2,2)/\allowbreak{}cc-pVDZ calculations using six CPU workers on an Intel Xeon Platinum 8480C system. Each alkene uses 4,000 common configurations from 20 complete trajectories, with three PEACE repeats after warm-up. Error bars and shading indicate 95\% confidence intervals from paired trajectory-block bootstrap resampling\cite{si16}. Ratios are computed from the mean wall times and are specific to these hardware and workflow settings.}{suppfig:14}\hfill\SIhalffigure{width=\linewidth}{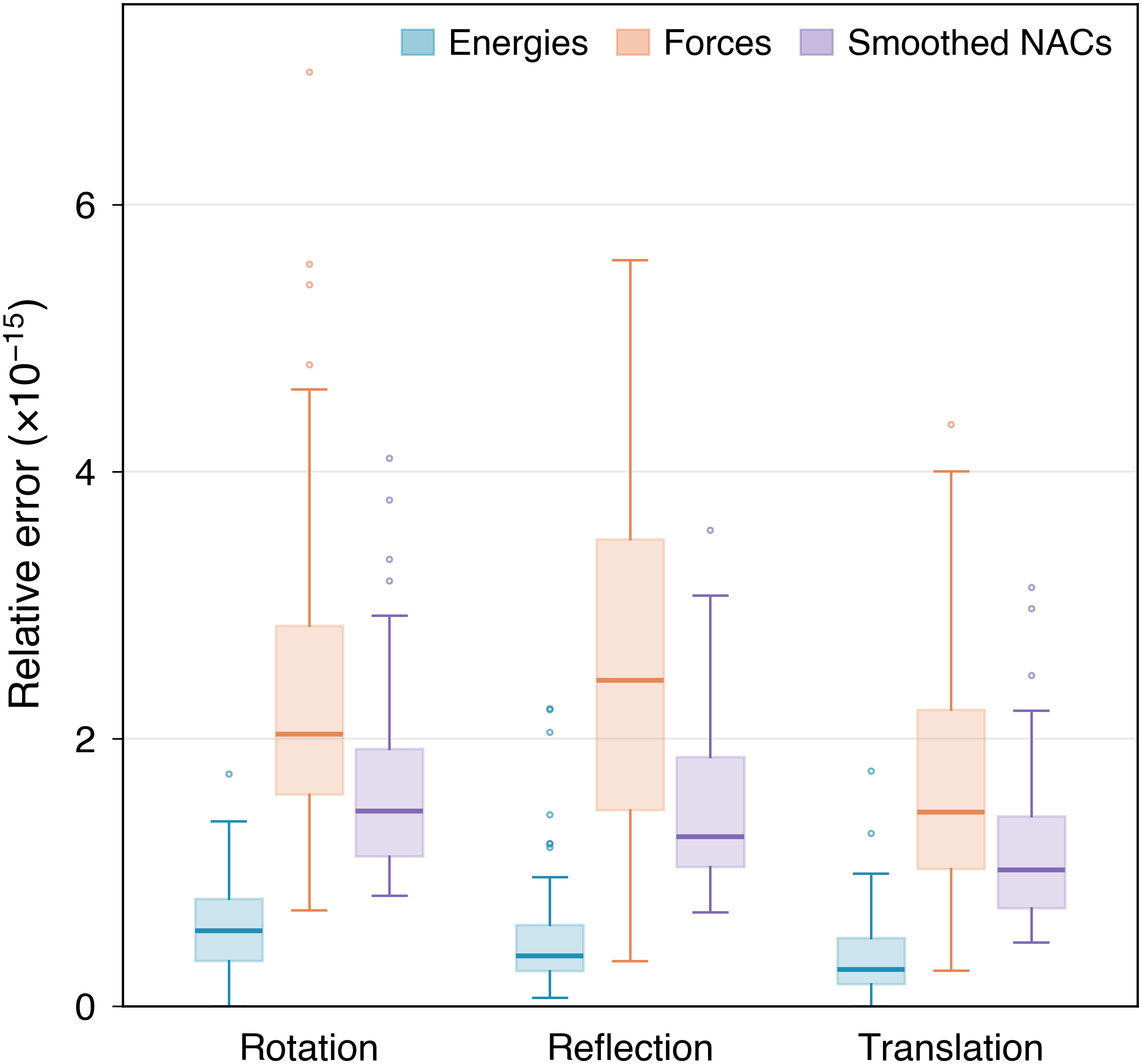}{\textbf{Numerical tests of symmetry consistency.} Relative errors in energies, forces, and smoothed nonadiabatic couplings (NACs) were evaluated in FP64 across 32 molecular geometries subjected to rotations, reflections, and translations. Boxes span the interquartile range, with horizontal lines indicating medians. Whiskers extend to the most extreme observations within 1.5 times the interquartile range; points denote outliers.}{suppfig:15}
\par\vspace{8pt}
\SItable{\textbf{Static mean absolute errors across four molecules.} Errors compare PEACE with SchNet- and PaiNN-based SPaiNN models\cite{si1}. NAC rows report raw derivative-coupling errors in atomic units.}{tab:mae}{\begin{tabular}{@{}llrrr@{}}
\toprule
Dataset & Quantity & SPaiNN (SchNet) & SPaiNN (PaiNN) & PEACE \\
\midrule
\multirow{3}{*}{C\textsubscript{2}H\textsubscript{4}} & Energies (eV) & 0.026 & 0.026 & \textbf{0.008} \\
 & Forces (eV/Å) & 0.146 & 0.113 & \textbf{0.036} \\
 & NACs (Bohr\textsuperscript{-1}) & 0.096 & 0.099 & \textbf{0.028} \\
\addlinespace[4pt]
\multirow{3}{*}{C\textsubscript{3}H\textsubscript{6}} & Energies (eV) & 0.031 & 0.016 & \textbf{0.013} \\
 & Forces (eV/Å) & 0.149 & 0.091 & \textbf{0.042} \\
 & NACs (Bohr\textsuperscript{-1}) & 0.081 & 0.044 & \textbf{0.016} \\
\addlinespace[4pt]
\multirow{3}{*}{C\textsubscript{4}H\textsubscript{8}} & Energies (eV) & 0.064 & 0.039 & \textbf{0.022} \\
 & Forces (eV/Å) & 0.172 & 0.105 & \textbf{0.054} \\
 & NACs (Bohr\textsuperscript{-1}) & 0.059 & 0.029 & \textbf{0.018} \\
\addlinespace[4pt]
\multirow{3}{*}{CH\textsubscript{2}NH\textsubscript{2}\textsuperscript{+}} & Energies (eV) & 0.099 & 0.063 & \textbf{0.026} \\
 & Forces (eV/Å) & 0.419 & 0.316 & \textbf{0.109} \\
 & NACs (Bohr\textsuperscript{-1}) & 0.180 & 0.140 & \textbf{0.053} \\
\bottomrule
\end{tabular}}
\par\vspace{8pt}
\SItable{\textbf{Dataset partitions for training and evaluation.}}{tab:splits}{\begin{tabular}{@{}lrrr@{}}
\toprule
Dataset & Training & Validation & Test \\
\midrule
C\textsubscript{2}H\textsubscript{4} & 6000 & 500 & 1000 \\
C\textsubscript{3}H\textsubscript{6} & 6000 & 500 & 1000 \\
C\textsubscript{4}H\textsubscript{8} & 12000 & 2000 & 1000 \\
CH\textsubscript{2}NH\textsubscript{2}\textsuperscript{+} & 2500 & 250 & 1250 \\
CH\textsubscript{2}S & 4000 & 200 & 503 \\
\bottomrule
\end{tabular}}
}

\vspace{7pt}

\begin{multicols}{2}

\renewcommand{\refname}{Supplementary References}

\end{multicols}

\fi

\end{document}